\documentclass[aps,prd,twocolumn,showpacs,preprintnumbers,nofootinbib,superscriptaddress,groupedaddress,10pt]{revtex4-1}
\usepackage{graphicx}
\usepackage{epsfig}
\usepackage{bm}
\usepackage{amssymb}
\usepackage{color}
\usepackage{float}
\usepackage{eufrak}
\usepackage{amsmath}
\usepackage{dcolumn}

\def\doi{http://doi.org}

\def\be{\begin{eqnarray}}
\def\ee{\end{eqnarray}}

\begin{document}

\title{Anomalous decay rate of quasinormal modes in Reissner-Nordstr\"om black holes}

\author{R. D. B. Fontana}
\email{rodrigo.fontana@uffs.edu.br}
\affiliation{Universidade Federal da Fronteira Sul, Campus Chapec\'o-SC Rodovia SC 484 - Km 02, Fronteira Sul, CEP 89815-899, Brasil}

\author{P. A. Gonz\'{a}lez}
\email{pablo.gonzalez@udp.cl} \affiliation{Facultad de
Ingenier\'{i}a y Ciencias, Universidad Diego Portales, Avenida Ej\'{e}rcito
Libertador 441, Casilla 298-V, Santiago, Chile.}

\author{Eleftherios Papantonopoulos}
\email{lpapa@central.ntua.gr} \affiliation{Physics Division,
National Technical University of Athens, 15780 Zografou Campus,
Athens, Greece.}

\author{Yerko V\'asquez}
\email{yvasquez@userena.cl}
\affiliation{Departamento de F\'isica y Astronom\'ia, Facultad de Ciencias, Universidad de La Serena,\\
Avenida Cisternas 1200, La Serena, Chile.}

\date{\today }

\begin{abstract}

The anomalous decay rate of the quasinormal modes occurs when the longest-lived modes are the ones with higher angular number.
Such behaviour has been recently studied in different static spacetimes, for scalar and fermionic perturbations, being observed in both cases. In this work, we extend the existent studies to the charged spacetimes, namely, the Reissner-Nordstr\"om, the Reissner-Nordstr\"om-de Sitter and the Reissner-Nordstr\"om-Anti-de Sitter black holes. We show that the anomalous decay rate behaviour of the scalar field perturbations is present for every charged geometry in the photon sphere modes, with the existence of a critical scalar field mass whenever $\Lambda \geq 0$. In general, this critical value of mass increases with the raise of the black hole charge, thus rendering a minimum in the Schwarzschild limit. We also study the dominant mode/family for the massless and massive scalar field in these geometries showing a non-trivial dominance of the spectra that depends on the black hole mass and charge.

\end{abstract}


\maketitle

\flushbottom


\section{Introduction}

The quasinormal modes (QNMs) and quasinormal frequencies (QNFs)
\cite{PhysRev.108.1063, Zerilli:1971wd, Kokkotas_1999, Nollert:1999ji, Konoplya_2011, Berti_2009} are powerful tools in studying the properties of black holes and compact objects at large distances. Recently the detection of gravitational waves \cite{PhysRevLett.116.061102} stimulated further the study of such vibrations that could emerge as the final stage of large coalescing objects. One main importance of the knowledge of QNMs and QNFs is the information they accommodate of a perturbed object, such as mass, charge and spin, along with its linear stability. Additionally, they provide a spectrum that is independent of the initial conditions of the perturbation and depends only on the fundamental constants of the system. Such spectrum of QNMs consists of complex frequencies, $\omega=\omega_R+i \omega_I$, in which the real part $\omega_R$ determines the oscillation frequency of the modes, while the complex part $\omega_I$ determines their exponential decaying timescale (for a review on QNM modes see \cite{Kokkotas_1999, Berti_2009}). Also, the QNMs represent solutions of the equation that dominates the time profile evolution during a limited interval of time, namely, the ringing phase. The outcome of the field propagation in a
 black hole geometry is a field profile divided into three distinct periods: the initial burst (considered as gauge, of free choice), coming from the perturbed wave put to evolve in the system, the quasinormal ringing phase and subsequent to it the late time behaviour that varies according to the spacetime analysed. There are several methods for calculating the QNFs, as the most well known Mashhoon method, Chandrasekhar-Detweiler method, Wentzel-Kramers-Brillouin(WKB) method, Frobenius method, method of continued fractions, (with the Nollert improvements), asymptotic iteration method (AIM) and improved AIM. For an extensive review, see \cite{Konoplya_2011}.

Different studies of the QNFs found that when a Schwarzschild or Kerr black hole is perturbed by a probe massless scalar field then the longest-lived modes are those with higher angular number, $\ell$. In a physical system  this behaviour can be understood from the fact that the more energetic modes with higher angular number $\ell$ would have faster decaying rates. However, if the perturbed probe scalar field is massive, then in the case of a light scalar field  the longest-lived QNMs are those with a high angular number $\ell$, whereas for a heavy scalar field the longest-lived modes are those with a low angular number $\ell$ \cite{Konoplya_2005, Konoplya_2006, Dolan_2007, Tattersall_2018}.  Such different behaviour occurs because in the regime of a small scalar mass, the fluctuations of the field can maintain the QNMs to live longer even if the angular number $\ell$ is large. This anomalous behaviour is characterized by a critical value of the mass of the scalar field, and the anomalous decay rate for small mass scale of the field was recently discussed in \cite{Lagos_2020}.

The propagation of scalar fields  in the Schwarzschild-dS and Schwarzschild-AdS black hole backgrounds was studied e. g. in \cite{Aragon:2020tvq}. There, the effect of the introduction of a cosmological constant on the anomalous decay of QNMs was investigated. The QNMs in the background of a Schwarzschild-dS black hole are characterized by the photon sphere family of modes (oscillatory modes) and by the dS modes family consisting of purely imaginary QNFs, for small scalar masses. For the photon sphere modes it was shown that the presence of the cosmological constant leads to the decrease of the real oscillation frequency and to a slower decay when the mass of the scalar field increases for a fixed angular harmonic number. Furthermore, it was shown  the existence of an anomalous decay rate of QNMs, i.e, the absolute values of the imaginary part of the QNFs diminishes when the angular harmonic numbers increases if the mass of the scalar field is smaller than a critical value. The effect of the cosmological constant in such cases is to shift the values of the critical masses i.e. when the cosmological constant increases the value of the critical mass also increases.

For the dS modes family it was shown that for a fixed value of the black hole mass, the purely imaginary QNFs can be dominant depending on the scalar field mass and the angular harmonic number. Also, a faster decay is observed when the $\ell$ parameter increases, as well as, when the scalar field mass increases up to the point where the QNFs acquire a real part. After that, the decay is stabilized and the frequency of the oscillations increases when the scalar field mass increases. Furthermore, it was found that this family does not present an anomalous behaviour of the QNFs, for the range of scalar field mass analysed. Further investigations also shown the anomalous decay rate behaviour for fermionic fields in the background of Schwarzschild de Sitter black holes \cite{Aragon:2020teq} and for accelerating black holes \cite{Destounis:2020pjk}.

In the case of a Schwarzschild-AdS black hole the decay rate of the QNMs of massive scalar fields present the anomalous behaviour. In this geometry, the effective potential at infinity always diverges, due to presence of the negative cosmological constant and consequently, the scalar field will probe such divergence. To avoid such mathematical divergences, the characteristic integration of fields in those AdS geometries considers a vanishing Cauchy surface at $\mathcal{I}$. In that case, the field also present a faster decay when its mass increases and when the angular harmonic numbers decrease.

The study of the anomalous decay of QNMs was extended to $f(R)$ modified gravity in \cite{NOJIRI_2007, Copeland:2006wr, Nojiri_2011, Clifton_2012, Capozziello_2007}. In \cite{Aragon:2020xtm} asymptotically de Sitter black holes solutions of $f(R)$ modified gravity were considered as background. The motivation of \cite{Aragon:2020xtm} was to investigate the competing effects of the cosmological term and the non-linear curvature terms on the behaviour of QNMs and QNFs generated by massless or massive scalar field perturbations. The black hole solution in the $f(R)$ gravity considered was characterized by a parameter $\beta$ which appears in the metric function of the solution. For  small values of this parameter the behavior of the solution is similar to that observed for Schwarzschild-dS black holes, i.e the anomalous  decay rate behaviour is present for small deviation of Schwarzschild-dS black holes. Thus, the deviation could be measured through the $\beta$ parameter indicating if the anomalous behaviour is present (small beta) or not (large beta) in the QNMs.

The structure of a charged black hole is richer than that of a chargeless spacetime, as long as it has more horizons which brings more types of the QNMs. In this work we will study the anomalous decay rate of QNMs for scalar field in the background of Reissner-Nordstr\"om (RN), Reissner-Nordstr\"om-de Sitter (RNdS)
and Reissner-Nordstr\"om-Anti-de Sitter (RNAdS) black holes, examining as well, the dominant mode/family for massless and massive scalar field in each spacetime.

In RN black holes, the ringing phase is characterized for a tower of the well-established photon sphere modes \cite{Cardoso_2009, Cardoso_2018, destounis2019strong}, and the last stage of evolution is that of a polynomial decay, coming from the Schwarzschild term of the potential \cite{Kokkotas_1999}. A secondary family of oscillations is also present in field evolution profiles (subdominant). That eventually takes control over the field propagation in the near-extremal limit of charge.

In the case of a RNdS geometry, on the other hand, we may identify more than two different families of QNMs, the dominant one taking control of the evolution in late times (what may be a purely imaginary decay, depending on the geometry and field initial parameters).

For RNAdS geometries, the secondary phase is again that of a quasinormal ringing phase, what, in such spacetimes, is represented by its oscillatory family of frequencies (not associated with the photon sphere orbit), as the black hole shows no cosmological horizon. The presence of a purely imaginary profile (near-extremal family of modes) is also present in the regime of high black hole charge. At late times, in the third stage of the field evolution, we may see a logarithmic decay \cite{Kehle_2019, kehle2018uniform}, the softest of all three black holes we here consider.

The work is organized as follows. In Sec. \ref{s3}, we study the backgrounds considered in the subsequent sections as well as the scalar perturbations. Then, in Sec. \ref{RNcase} we compute the QNMs for scalar fields in a RN black hole and we study the anomalous behaviour of the decay rate, analysing the dominant mode/family for massless and massive scalar field. In Sec. \ref{RNdScase}, we carry out a similar investigation for RNdS black hole. Then, in Sec. \ref{WKBJ} we present a semi-analytical analysis via the WKB method and in Sec. \ref{RNAdScase}, we consider RNAdS black holes as a background. We present our final remarks in Sec. \ref{conclusion}.

\section{Scalar perturbations in a charged black hole background}
\label{s3}

In this work we will consider a  probe scalar field in the background of a RN black hole with or without cosmological constant. In the case $\Lambda \neq 0$ we have the well-known RNdS - $\Lambda > 0$ -  or RNAdS - $\Lambda < 0$ - geometries. For all three geometries, we can write the metric in spherical coordinates as
\be
\label{eb1}
ds^2 = -fdt^2 + f^{-1}dr^2 + r^2d\Omega^2~,
\ee
in which $f=1-\frac{2M}{r} + \frac{Q^2}{r^2} -\frac{\Lambda}{3} r^2$ and $d\Omega^2 $ is the unitary 2-sphere line element. The causal structure is settled by the zeros of $f$, defining the respective null surfaces, the Cauchy ($r_C$), the event ($r_H$) and the cosmological ($r_\Lambda$) horizons, the last one only for RNdS.

We consider  the action,
\be
\nonumber
S=\int d^4x(S_b+S_p) = \int d^4x\Bigg( \left( \mathcal{R} - \frac{\Lambda}{2} - \frac{F^2}{4}\right) \\
\label{eb2}
 - \left(\frac{1}{2}g^{\mu \nu}\partial_\mu \Phi \partial_\nu \Phi  + \frac{1}{2}\mu^2 \Phi^2\right) \Bigg)~,
\ee
in which the first term generates the background represented by the solution (\ref{eb1}) and the second term represents a dynamical scalar field as studied in \cite{Molina_2004}, thus a perturbative term expected to decay in time, so that the geometry remain unchanged.

The Klein-Gordon motion equation is obtained through the variation of Eq. \ref{eb2} with respect to the scalar field and can be written as
\be
\label{eb5}
\frac{1}{\sqrt{-q}}\partial_\mu ( \sqrt{-g} \mathfrak{g}^{\mu \nu} \partial_\nu \Phi )  - \mu^2 \Phi =0~.
\ee
Now, by means of the following ansatz
\begin{equation}
\Phi =e^{-i\omega t} R(r) Y(\Omega) \,,\label{wave}
\end{equation}%
the equation reduces to
\begin{equation}
\frac{1}{r^2}\frac{d}{dr}\left(r^2 f(r)\frac{dR}{dr}\right)+\left(\frac{\omega^2}{f(r)}+\frac{\kappa}{r^2}-\mu^{2}\right) R(r)=0\,, \label{radial}
\end{equation}%
where we defined $\kappa=-\ell (\ell+1)$, with $\ell=0,1,2,...$ representing the eigenvalue of the Laplacian on the two-sphere and $\ell$ is the multipole number or the angular momentum of the field. Now, defining $R(r)=\frac{F(r)}{r}$ and the tortoise coordinate
$dr^*=\frac{dr}{f(r)}$, the wave equation can be written as a one-dimensional Schr\"{o}dinger-like equation
 \begin{equation}\label{ggg}
 \frac{d^{2}F(r^*)}{dr^{*2}}-V_{eff}(r)F(r^*)=-\omega^{2}F(r^*)\,,
 \end{equation}
 with an effective potential $V_{eff}(r)$ given  by
  \begin{equation}\label{pot}
 V_{eff}(r)=-\frac{f(r)}{r^2} \left(\kappa -  \mu^2 r^2-f^\prime(r)r\right)~,
 \end{equation}
whose behaviour at infinity is
\begin{equation}
    V_{eff}(r\rightarrow \infty)=-\frac{1}{9} r^2 \left(3 \Lambda  \mu^2-2 \Lambda ^2\right)\,~.
\end{equation}

In what follows we analyse the quasinormal spectra for the massive scalar field in each black hole background focusing in the study of the dominant mode/family with different values of scalar field masses. We verify that the presence of an anomalous behaviour relative to the dominance of different angular momenta in the spectra  occurs just in the case of the photon sphere (oscillatory) modes, preventing other solutions of such property.

\section{Reissner-Nordstr\"om quasinormal modes}
\label{RNcase}

The Reissner-Nordstr\"om black hole presents a spherically symmetrical solution without cosmological constant, but with mass and charge as geometrical properties. The existence of two horizons brings as a result of the scalar field perturbations two different families of quasinormal modes: the traditional photon sphere family (PS), associated with geodesical quantities in the eikonal limit and the near-extremal family (NE).

The existence of oscillatory and purely-imaginary transient scenarios in the field evolution, (for chosen black hole parameters) brings up the different behaviours of such families of frequencies: the first quasinormal family, well-described along the last four decades in the literature, is the one usually exhibited in field evolution profiles, after a initial burst of perturbation. It represents the oscillatory 'brand' associated with the existence of a transient behaviour for the partial differential equation (PDE) of the field.

On the other hand, a second family of modes, describing an independent solution to the scalar equation was firstly described in \cite{hod2017quasibound}, nominated {\it quasi-bound states}. In such case, a chargeless scalar field has a purely-imaginary spectrum, proportional to the surface gravity of black hole event (or Cauchy) horizon. The dominance of such quasi-bound states emerges when the Cauchy and event horizons are very near to each other. Otherwise, in the regime $Q \ll M$, the PS modes take control of the field propagation (though the NE family is also present, subdominant). We analyse both families in the sequence.

\subsection{The photon sphere modes}

The oscillatory solutions can be acquired through the WKB method, described in \cite{Iyer:1986np,Iyer:1986nq,Kokkotas_1999,Kokkotas:1988fm,Seidel:1989bp,Konoplya_2003}. Alternatively they can also be obtained via integration in double null coordinates. The spectrum appears in the secondary stage of the field evolution - being the signal captured through the prony method. As long as the convergence of both calculations is very good for the RN black hole, we choose to obtain the frequencies with the WKB method.

\begin{figure*}[!t]
\begin{center}
\includegraphics[width=0.40\textwidth]{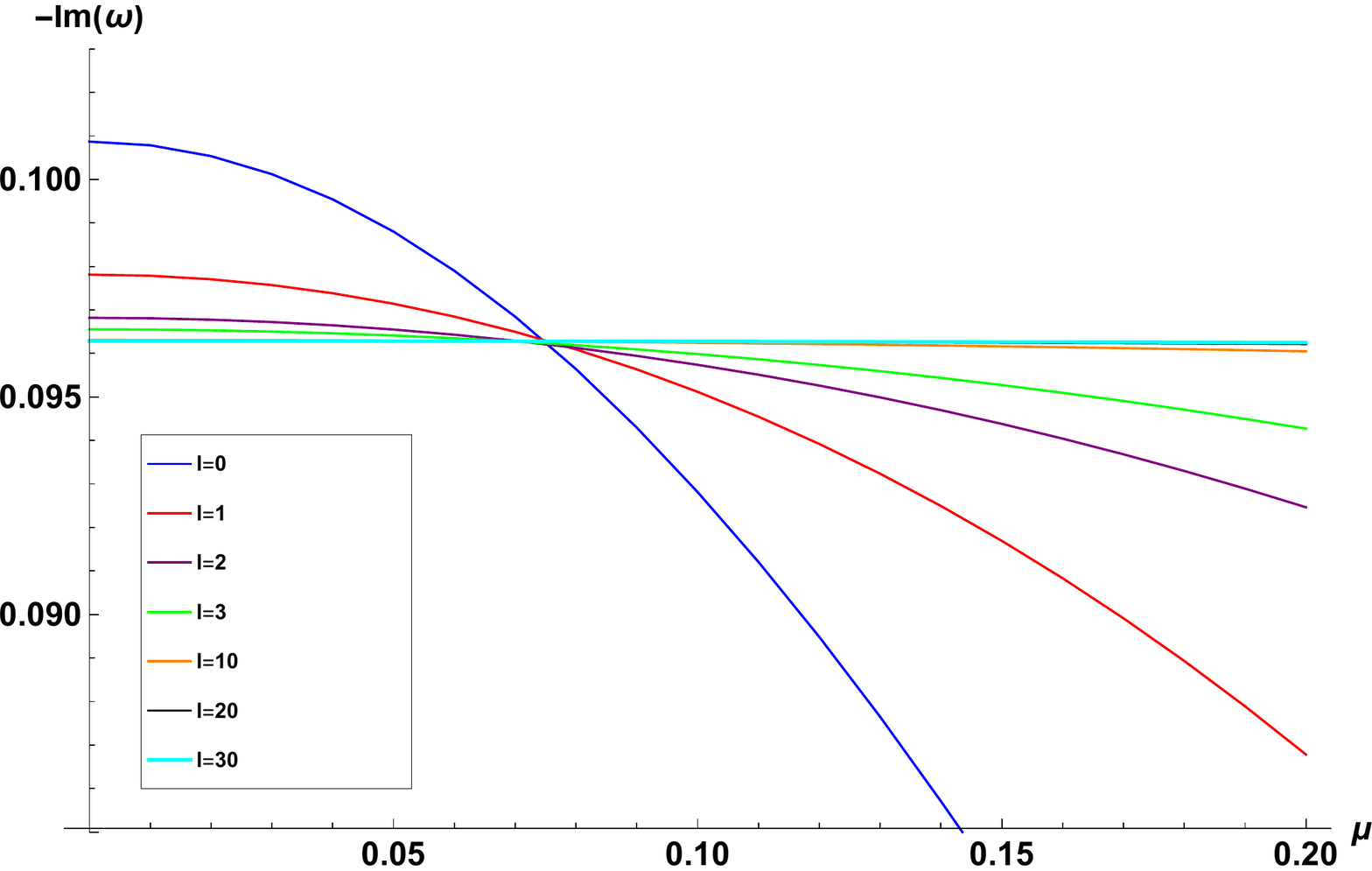}
\includegraphics[width=0.40\textwidth]{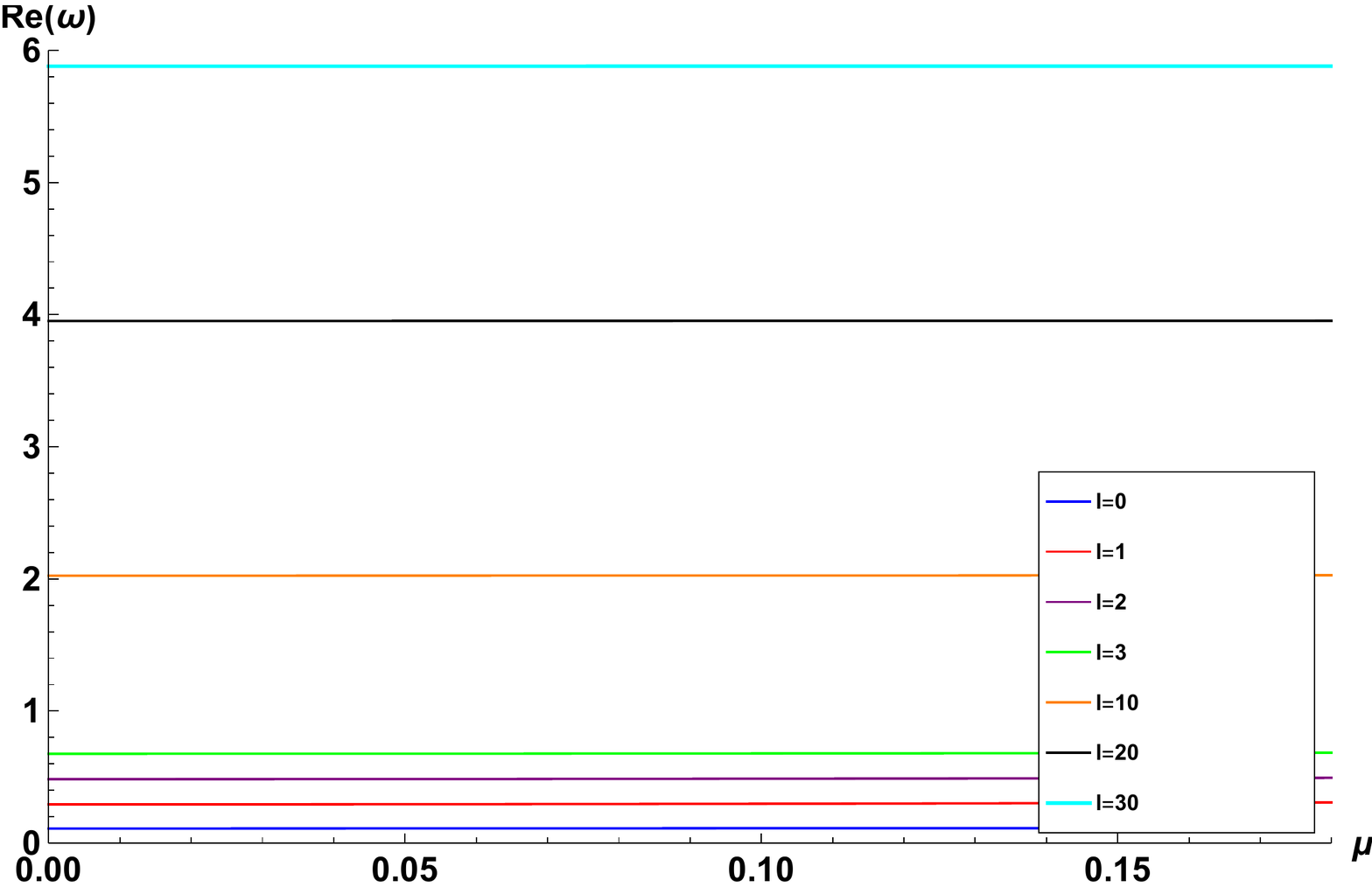}
\end{center}
\vspace{-2.5cm}
\caption{The fundamental oscillatory quasinormal modes (PS) of a massive scalar field in a RN black hole with $M = Q/10 = 1$.}
\vspace{-2.0cm}
\label{rna1}
\end{figure*}
\begin{figure*}[!t]
\begin{center}
\includegraphics[width=0.40\textwidth]{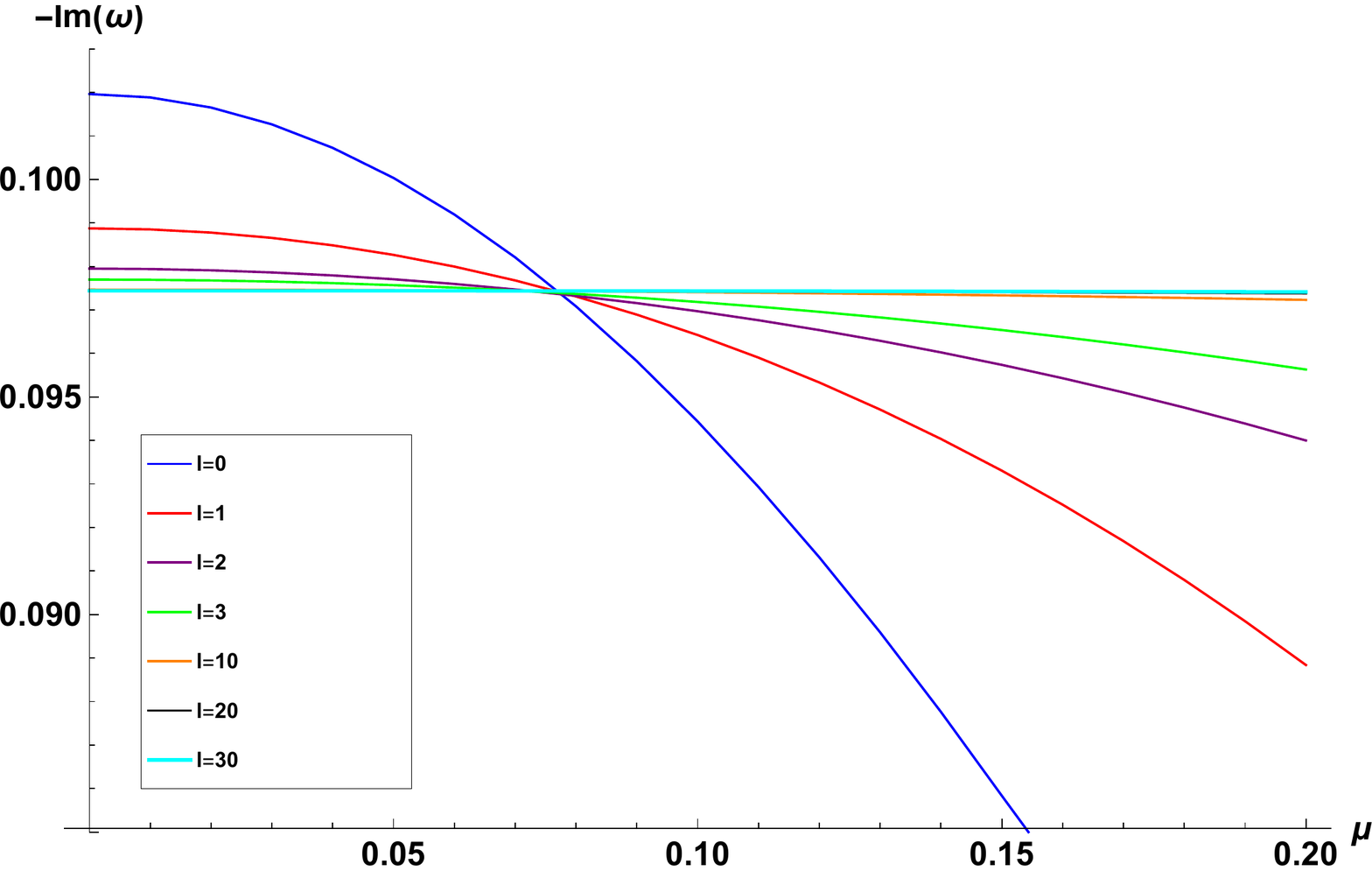}
\includegraphics[width=0.40\textwidth]{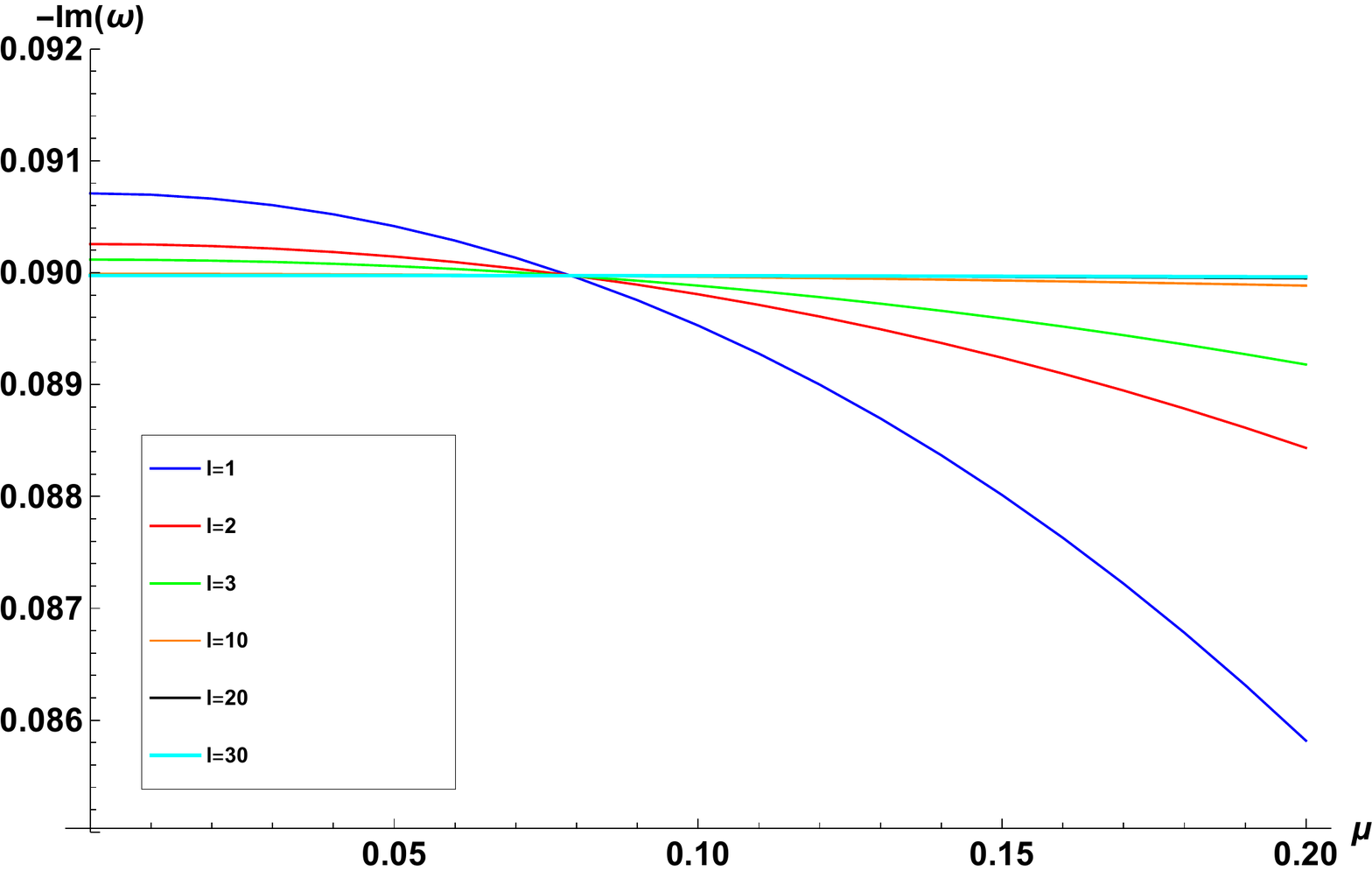}
\end{center}
\vspace{-2.5cm}
\caption{The oscillatory quasinormal modes of massive scalar field in a RN black hole with $M = 1$, $Q=1/2$ (left) and $Q=0.99$ (right).}
\label{rna2}
\end{figure*}

In Fig. \ref{rna1} we plot the quasinormal modes - both real (right) and imaginary (left) parts - with different values of the scalar mass.  We can observe the anomalous decay rate of the quasinormal modes, where the longest-lived modes are the ones with higher angular number, as well as, the existence of a critical scalar field mass, i.e the graphic point $\mu_c$ in which the decay rate does not depend appreciably on the angular number, where, beyond this value the behaviour of the decay rate is inverted. Clearly, there is a transition of dominance relative to the scalar mass: up to $\mu \sim 0.08$, the lowest lying mode has $\ell=0$, and for $\mu \gtrsim 0.08$ it is represented by the frequency of the eikonal limit, $\ell \rightarrow \infty$. The real part of the QNM, otherwise, is proportional to the angular momentum of the field, as usually in RN geometries.
\begin{figure*}[!t]
\begin{center}
\includegraphics[width=0.3\textwidth]{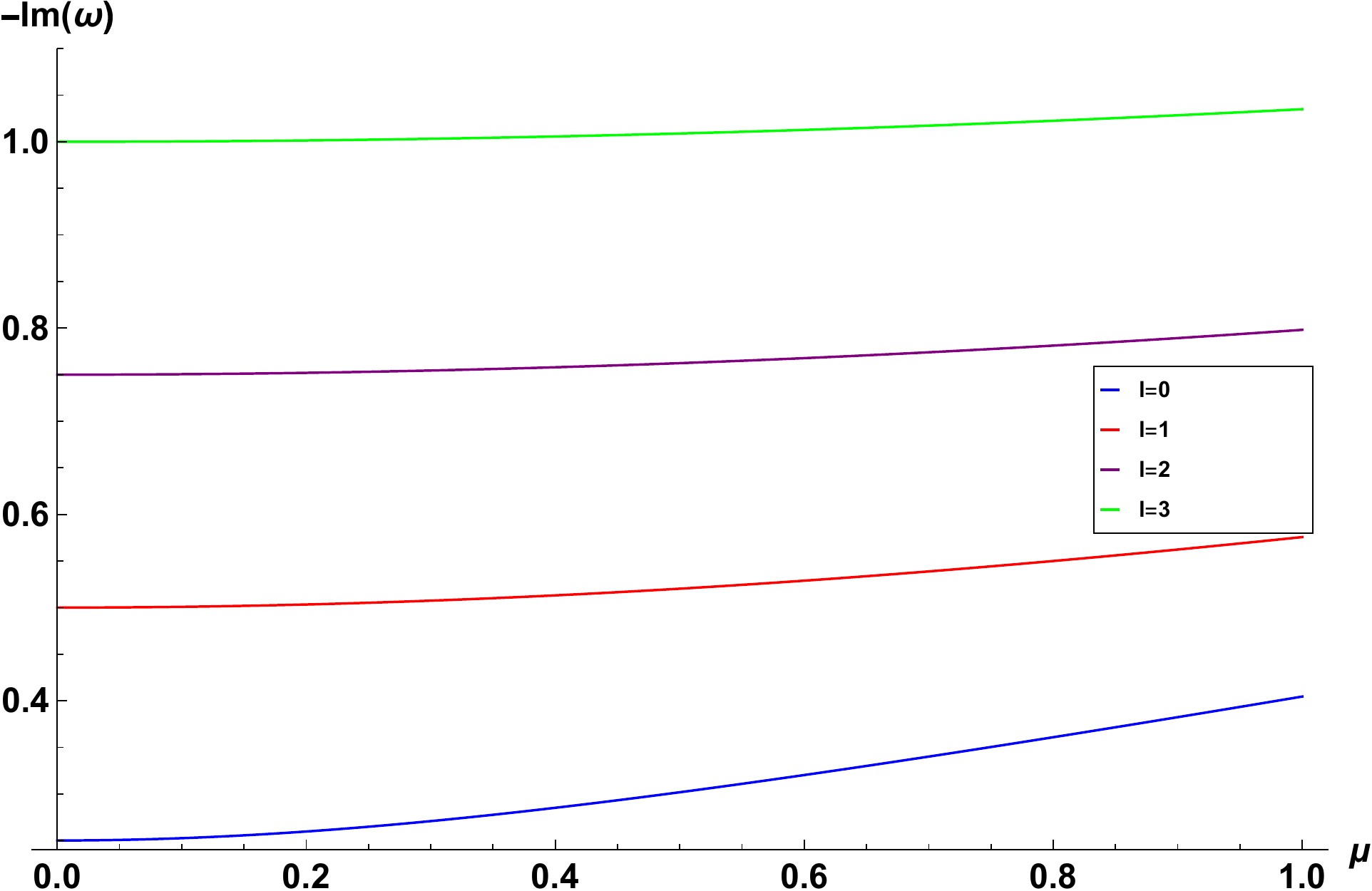}
\includegraphics[width=0.3\textwidth]{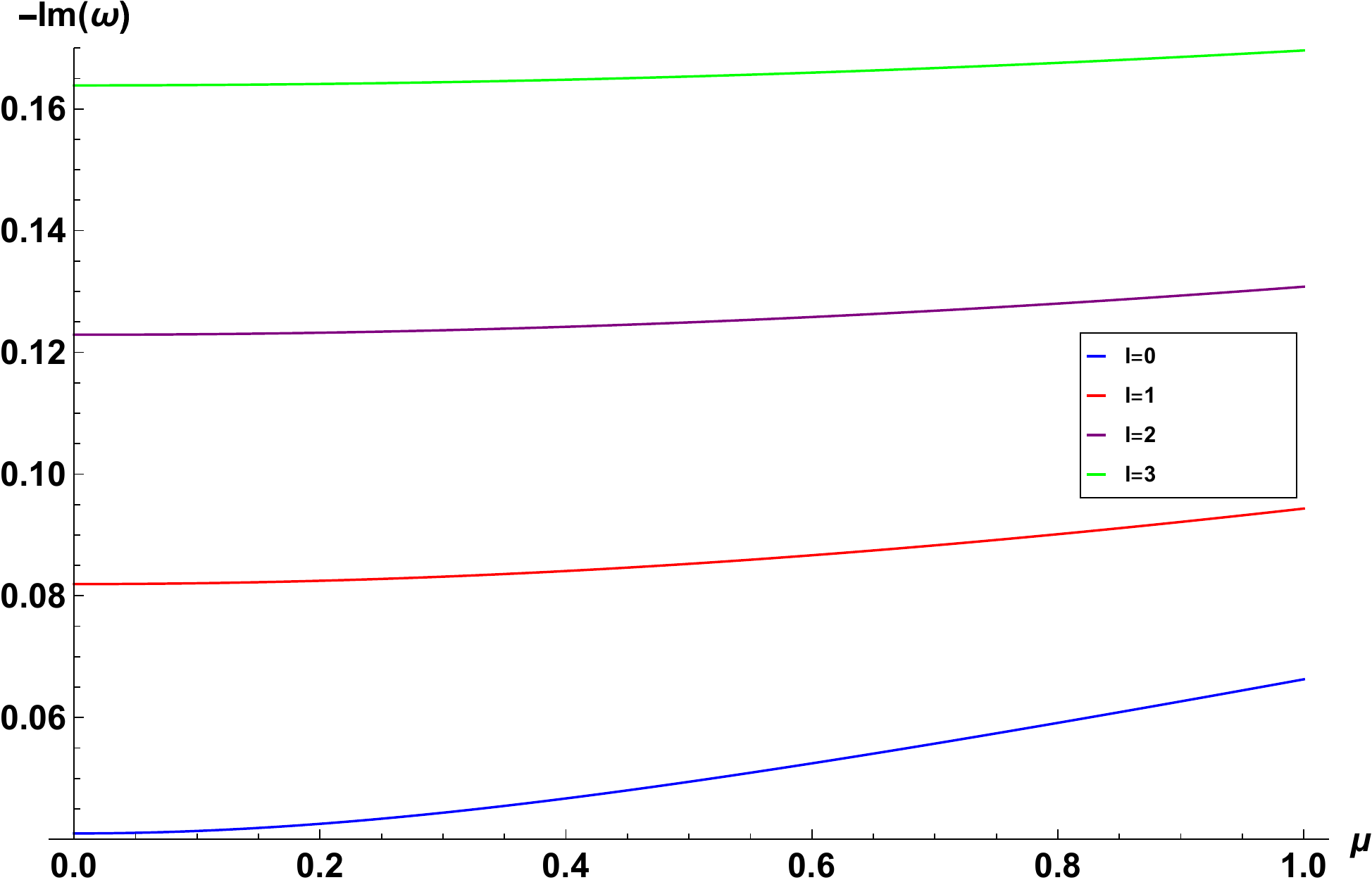}
\includegraphics[width=0.3\textwidth]{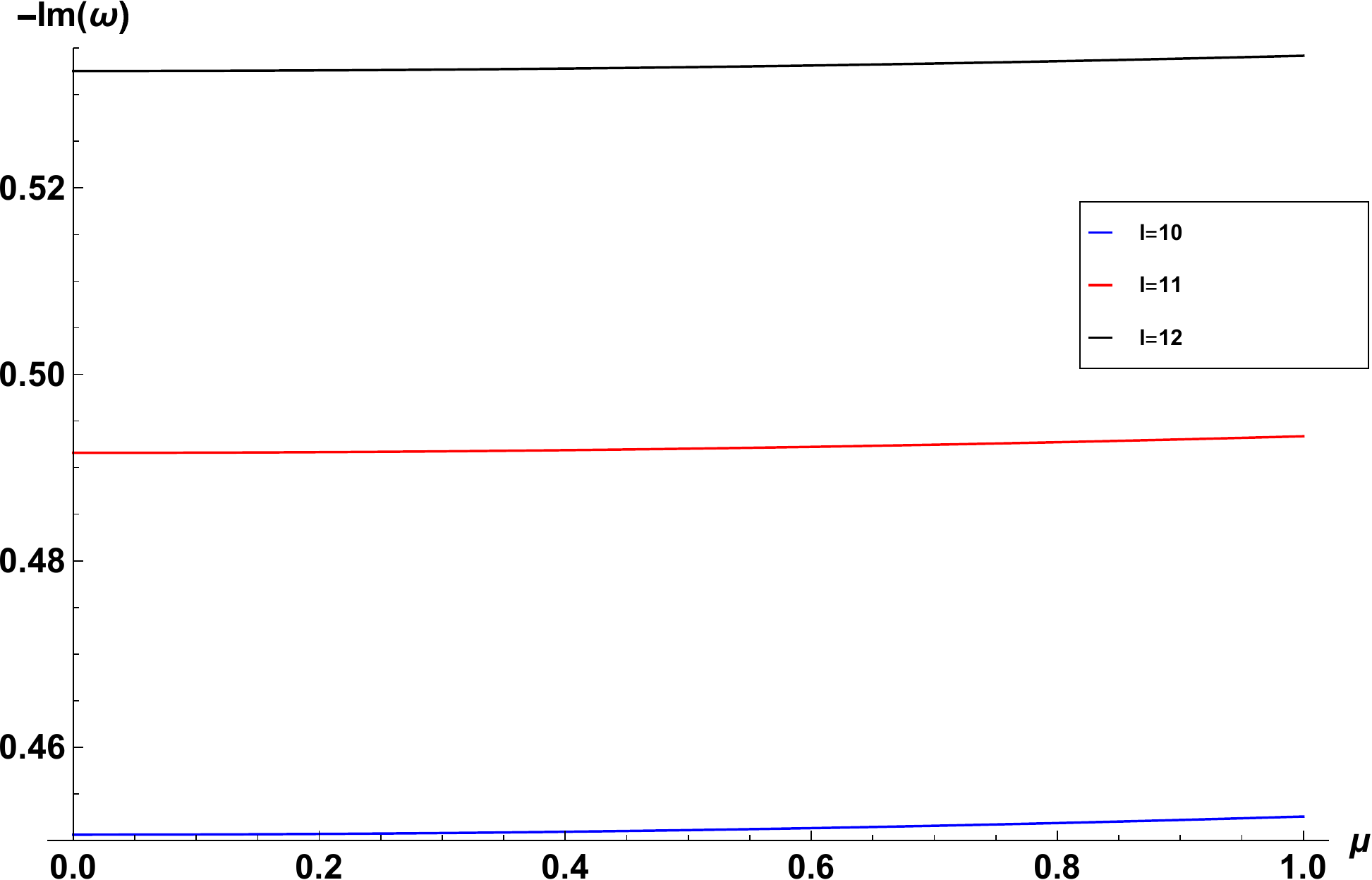}
\end{center}
\vspace{-0.5cm}
\caption{The near extremal purely imaginary quasinormal modes of a massive scalar field with different angular momenta and scalar field mass.  The geometry parameters read $M =1$, $Q=0.1$ (left)  and $Q=0.999$ (middle/right).}
\label{rnne1}
\end{figure*}
\begin{figure*}[!t]
\begin{center}
\includegraphics[width=0.40\textwidth]{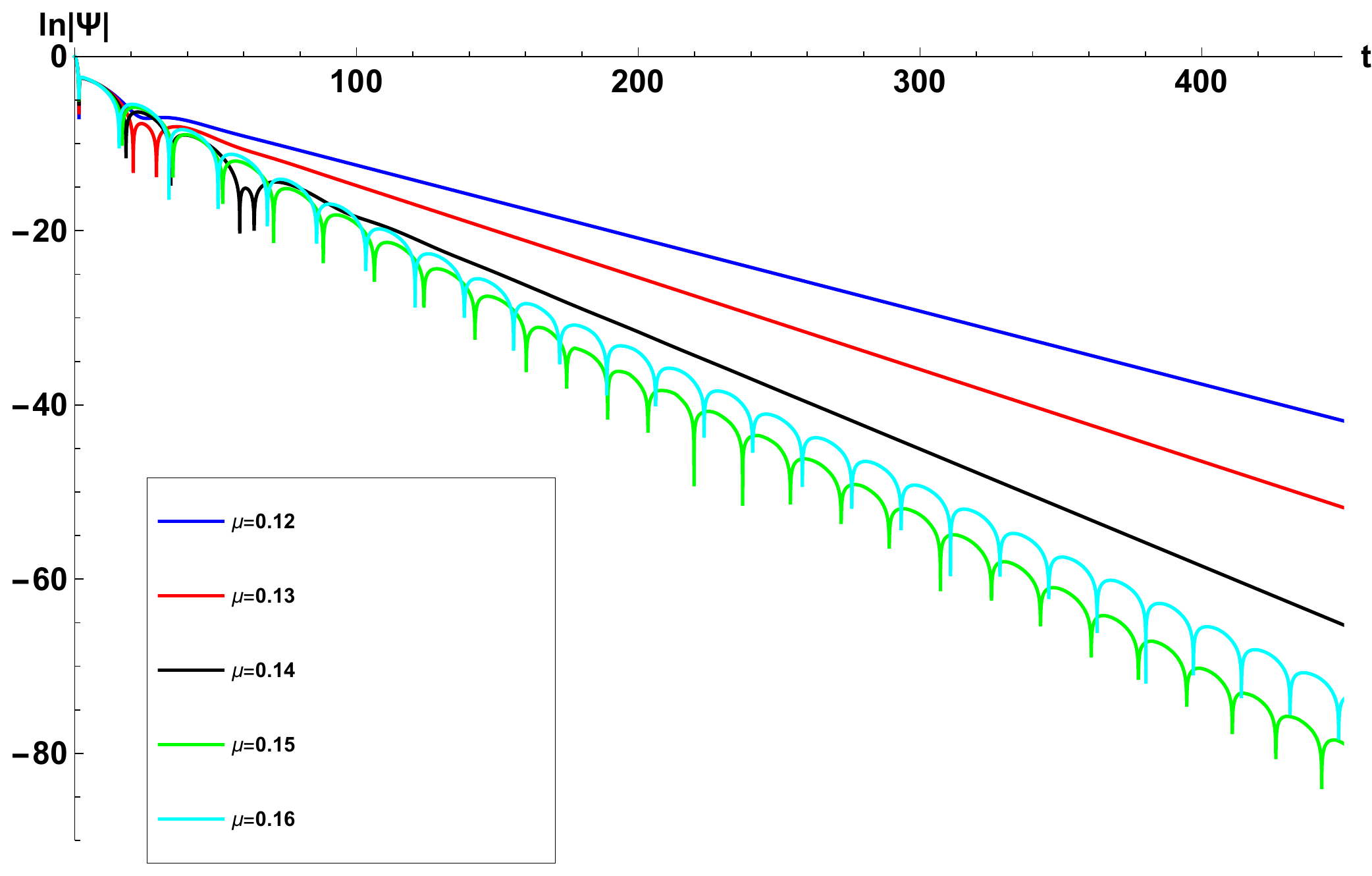}
\includegraphics[width=0.40\textwidth]{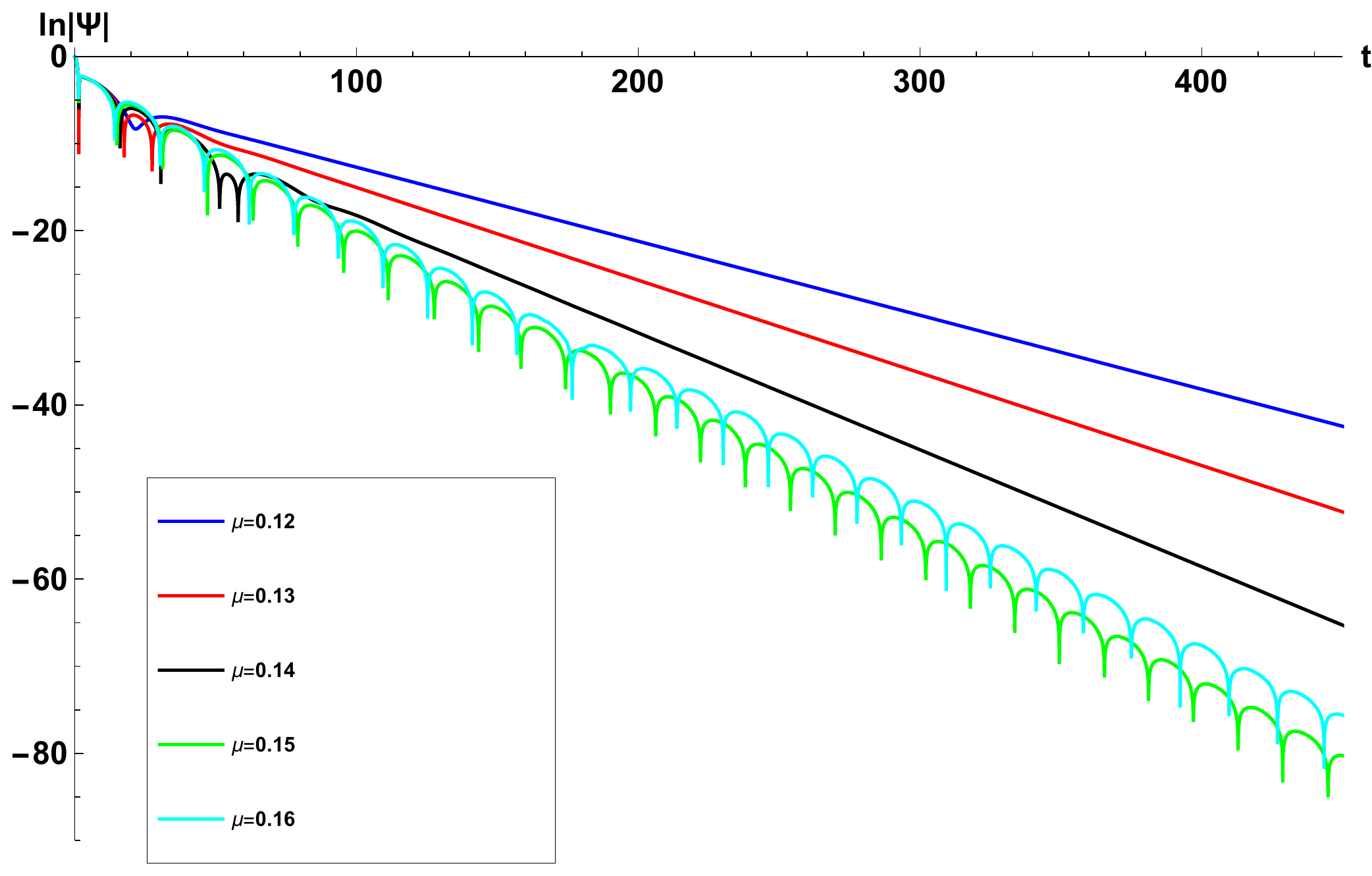}
\end{center}
\caption{The behaviour of the cosmological quasinormal modes in a RNdS black hole with $M=25 \Lambda = 1$, $\ell =0$, $Q=1/10$ (left) and $Q=6/10$ (right). The transition from a purely imaginary to an oscillatory profile occurs at $\mu \sim 0.145$, while for a pure de Sitter spacetime, according to (\ref{w1}) takes place at  $\mu \sim 0.173$.}
\label{fr1}
\end{figure*}

The same type of behaviour is obtained for different black hole parameters as we increase the value of $Q$. In Fig. \ref{rna2} we display the imaginary part of the photon sphere frequencies for higher values of charge. We can observe that the critical scalar field mass increases when the charge of the black hole increases.

\subsection{The near extremal modes}

A distinctive portrayal is obtained by the purely imaginary family whose characterization is dictated by
\be
\label{rnf1}
\omega = -i k_h \left(n+\frac{1}{2}+\sqrt{(\ell +1/2)^2 + M^2 \mu^2} \right)~,
\ee
in which $k_h$ is the surface gravity of the black hole, $n$ the overtone number, $\ell$ the angular momentum of the field and $\mu$ its mass. The presence of the near-extremal modes in the spectra is noticed when $\delta \equiv \frac{M-Q}{M}\ll 1$: since the lowest-lying frequency is given directly by the gravitational surface of the event horizon, the purely-imaginary solutions will be small in comparison to the PS solution (consequently dominant) whenever the temperature of the black hole is small. As a matter of comparison, the typical damping for Reissner-Nordström black hole of a PS mode is of order $\omega_I^{PS} \sim 0.1$ and this is achieved by the near-extremal family when $Q = 0.99M$. For smaller charges, we have higher damping factors in the NE family, e. g. $Q = 0.9$ and $0.8$ bring purely imaginary modes, $\omega_I^{NE} \sim -0.211i$ and $\omega_I^{NE} \sim -0.234i$, much higher than the fundamental PS quasinormal mode imaginary part. On the other hand, for very high values of black hole charges such NE solutions may not be ignored: $\delta = 10^{-3}$, $10^{-4}$ and $10^{-5}$ brings $\omega_I^{NE}=-0.0410i$, $\omega_I^{NE}=-0.0138i$ and $\omega_I^{NE}=-0.00447i$, regime in which this family takes control over the field evolution.
\begin{figure*}
\begin{center}
\includegraphics[width=0.40\textwidth]{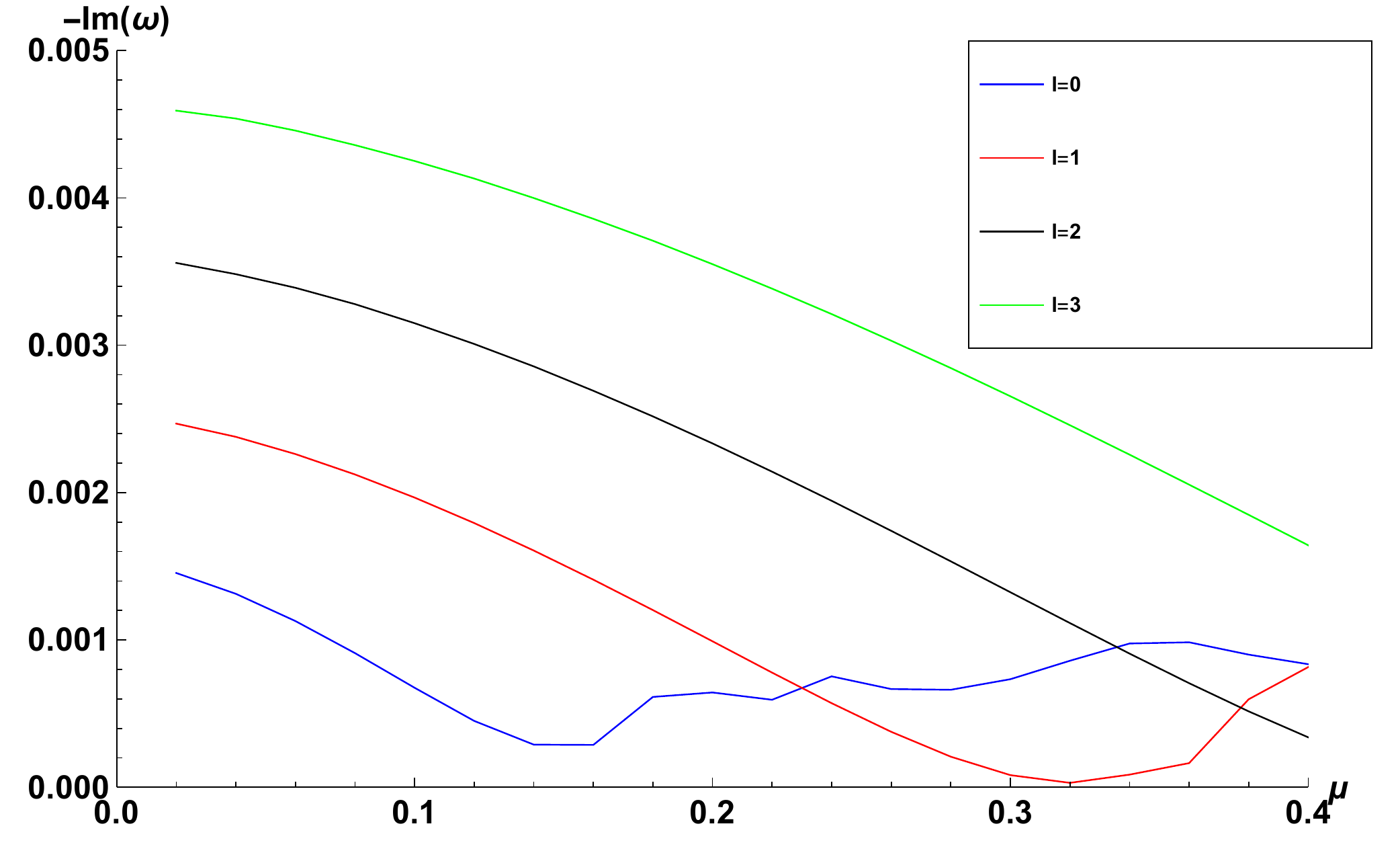}
\includegraphics[width=0.40\textwidth]{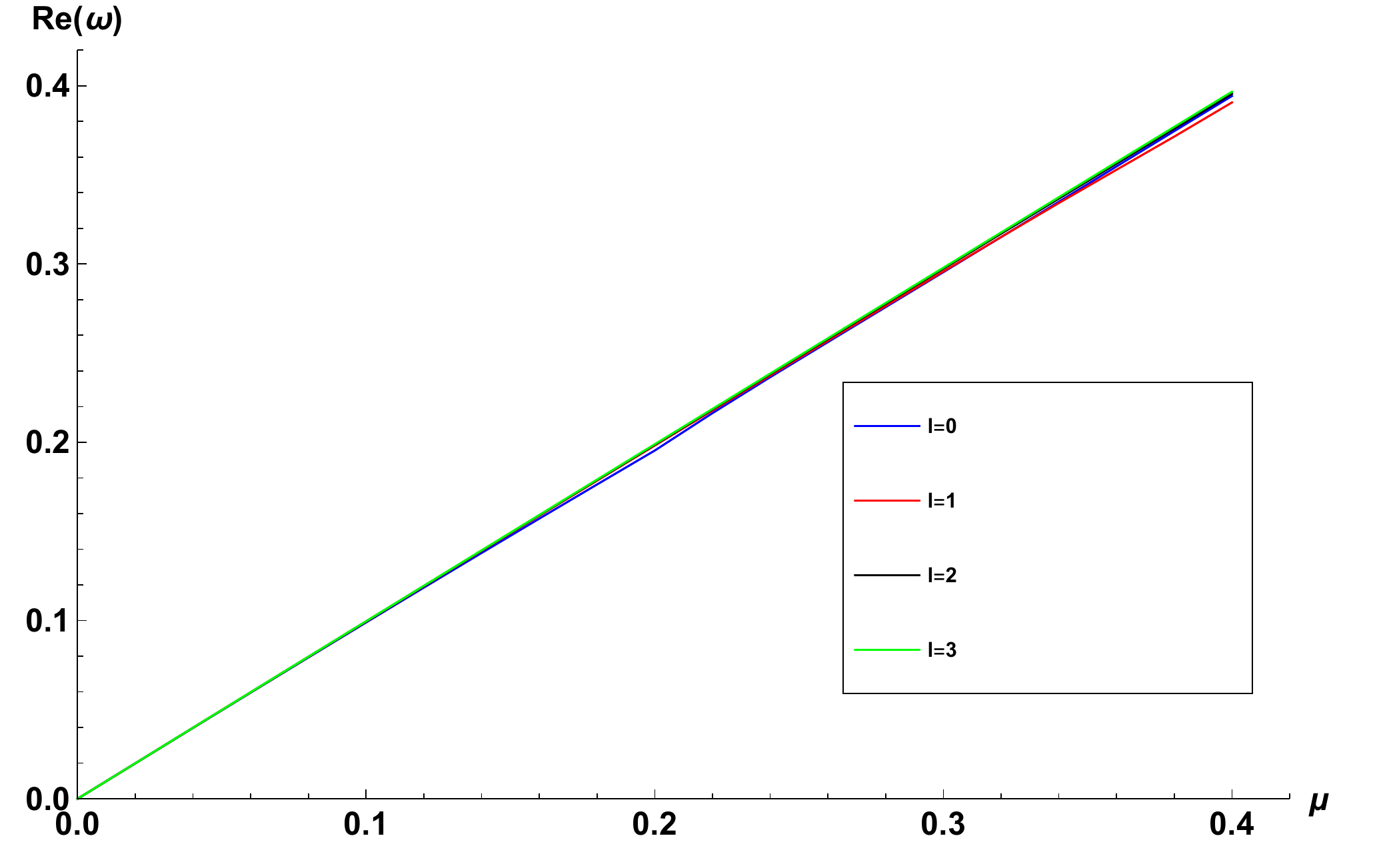}
\end{center}
\caption{The behaviour of the cosmological quasinormal modes in a RNdS black hole with $M = 2Q = 3.10^6 \Lambda = 1$.}
\label{fr2}
\end{figure*}

\begin{figure*}
\begin{center}
\includegraphics[width=0.4\textwidth]{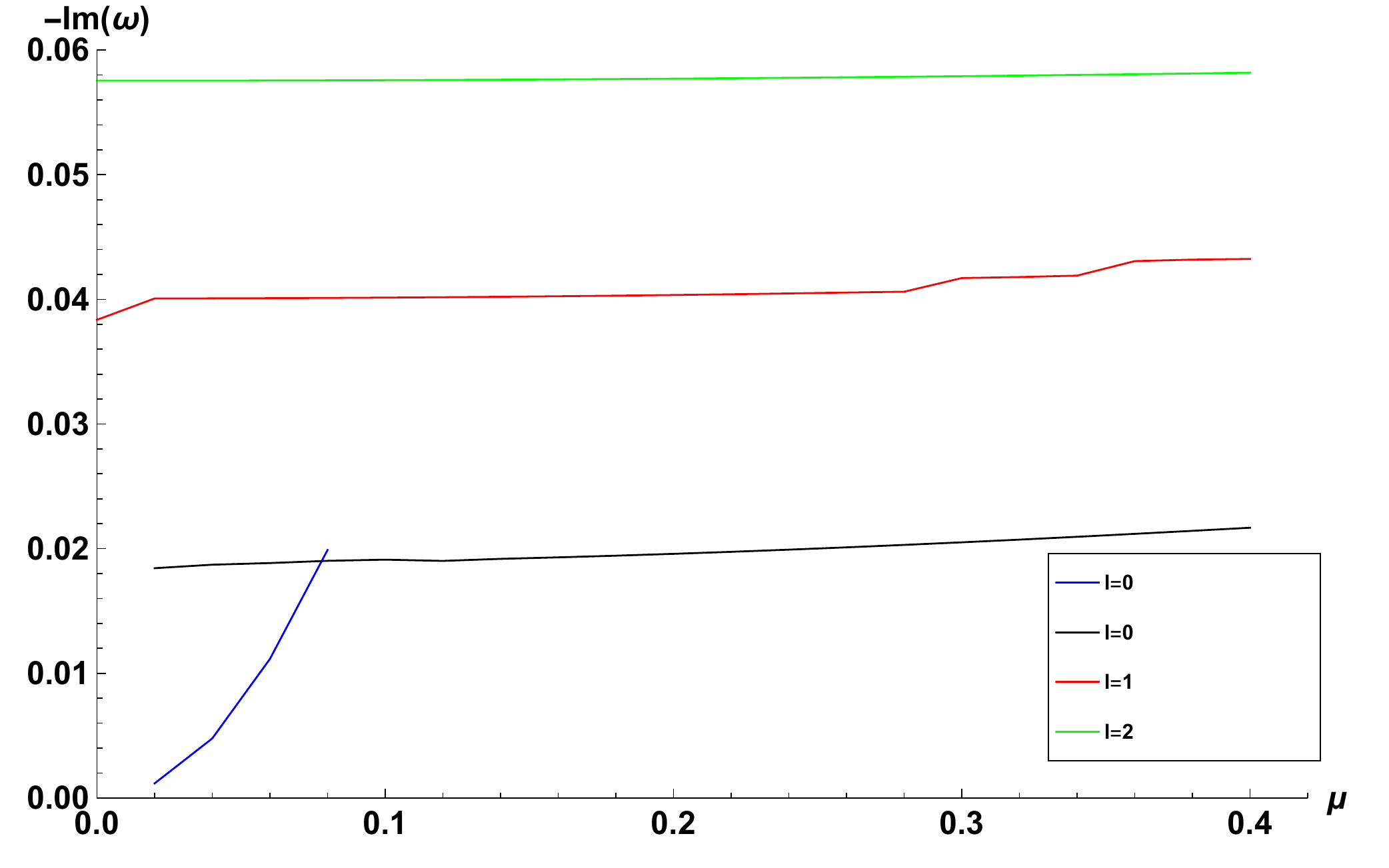}
\includegraphics[width=0.4\textwidth]{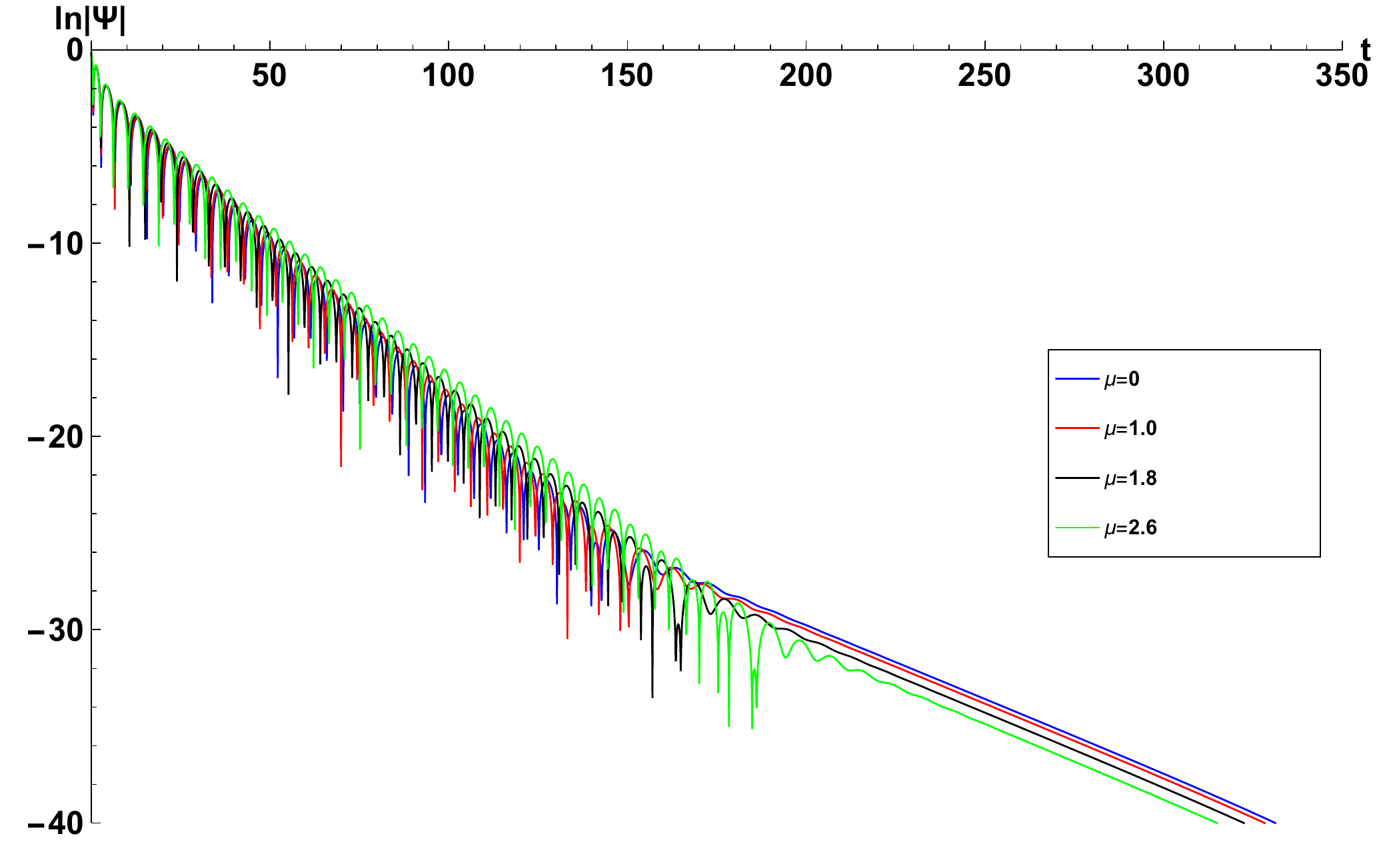}
\end{center}
\caption{The purely imaginary near extremal modes for a massive scalar field in RNdS (left) and field profile examples (right). The geometry parameters read
$M = Q/(0.9998Q_{max}) = 100\Lambda /3 = 1$ (left) and the field angular momentum $\ell = 1$ (right). The blue line (left pannel) stands for the QNMs in which the cosmological family of modes is dominant for the example (when $\ell =0$, until $\mu \sim 0.08$). All other modes represent dominance of near extremal frequencies.}
\label{fr4}
\end{figure*}
It is worth mentioning that the behaviour of the near-extremal frequencies is peculiar: the anomaly found in the PS case (transitional behaviour of dominance), is absent here; increasing the scalar field mass does not produce an interplay in the dominant modes (high to small angular momentum): the growth of $\omega_I$ is proportional to the augmentation of both properties of the field, what can be seen in Fig. \ref{rnne1}.

The main  conclusion from such analysis is that the anomalous behaviour for the RN black hole is present  away from the near-extremal regime: whenever $\delta \lesssim 1$, the PS modes are in control of the field evolution. In this case, for small masses of scalar field (or zero), $\ell =0$ is the dominant mode, while for high masses $\mu \gtrsim \mu_c$, the eikonal modes take control. On the other way, if $\delta \lesssim 10^{-3}$, the dominant mode has $\ell = 0$, for every scalar field mass.

The existence of interplay in the dominance (small to high $\ell$) exhibited only in one family of oscillations is demonstrated to be the case also for the de Sitter spacetime. We study the issue in the next section, by examining the correspondent black hole (that possesses three different null surfaces $f=0$), doing a thorough analysis inside each regime of dominance.

\section{Reissner-Nordstr\"om de Sitter quasinormal modes}
\label{RNdScase}

The Reissner-Nordstr\"om de Sitter black hole has three different horizons specified in section \ref{s3}, correspondingly having three known families of QNFs. Each family of modes takes control over the field evolution in different ranges of geometric parameters, as we may further describe.

We investigate the existence of interplay in the dominance of field evolution, for each solution, with two different methods. One of them is the integration in double-null coordinates, together with the prony method - both techniques used for obtaining the QNMs when the charge is very near to extremality, or when the cosmological constant is very small, cases in which the WKB modes are sub-dominant. The WKB frequencies are also present (and checked with the following method) in the first part of the field evolution. The other one is the WKB approximation - used for parameters in which the field evolution oscillates perpetually, that is, the field profile is not taken by other modes, mainly purely imaginary.

\subsection{The cosmological modes}

Associated with the presence of a cosmological constant (and horizon), these modes resembled those of a pure de Sitter spacetime. They were firstly described in \cite{Du_2004}. The waves may have a purely imaginary evolution or an oscillatory profile, if the mass surpasses a critical value \cite{Aragon:2020tvq}. In a pure dS geometry, the resonances are determined by two branches \cite{Fontana_2019}, from which the most important in our investigation is the lowest lying solution, written as
\be
\label{w1}
\omega = -i\sqrt{\frac{\Lambda}{3}}\left(2n  + \ell + \frac{3}{2} \pm \sqrt{\frac{9}{4}-\frac{3\mu^2}{\Lambda }} \right)~.
\ee

In the above relation, the critical mass for which an oscillatory $\omega$ emerges is $\mu = \sqrt{1.5\Lambda}$. However such critical limit is
lower for RNdS black hole, which can be seen in Fig. \ref{fr1} (for smaller charges a higher deviation occurs). The massless cosmological modes are dominant when the cosmological constant is small. In that case, expression (\ref{w1}) for the fundamental mode, $n=0$ and $\ell=1$  is smaller then the imaginary part of the photon sphere modes as well as the near extremal frequencies\footnote{The near extremal family appears to be relevant only when the charge is significantly near its extremal value, similar to what is found in the above section.}.

The results obtained for the cosmological modes suggest an interesting fact: there is no interchange in dominant modes going from small to high scalar field mass in relation to $\ell$, i. e., the absence of such anomalous behaviour. This is the case as it can be observed in Fig. \ref{fr2}. Although the imaginary part oscillates for higher values of field mass, no interplay between the eikonal and the zero angular momentum modes occurs relative to the dominance in the spectrum.

The interesting fact is that the real part behaves exactly as in a pure de Sitter spacetime, dictated by (\ref{w1}),
\be
\label{wa2}
Re( \omega ) \sim \mu~,
\ee
for all $\ell$.

\begin{figure}
\begin{center}
\includegraphics[width=0.4\textwidth]{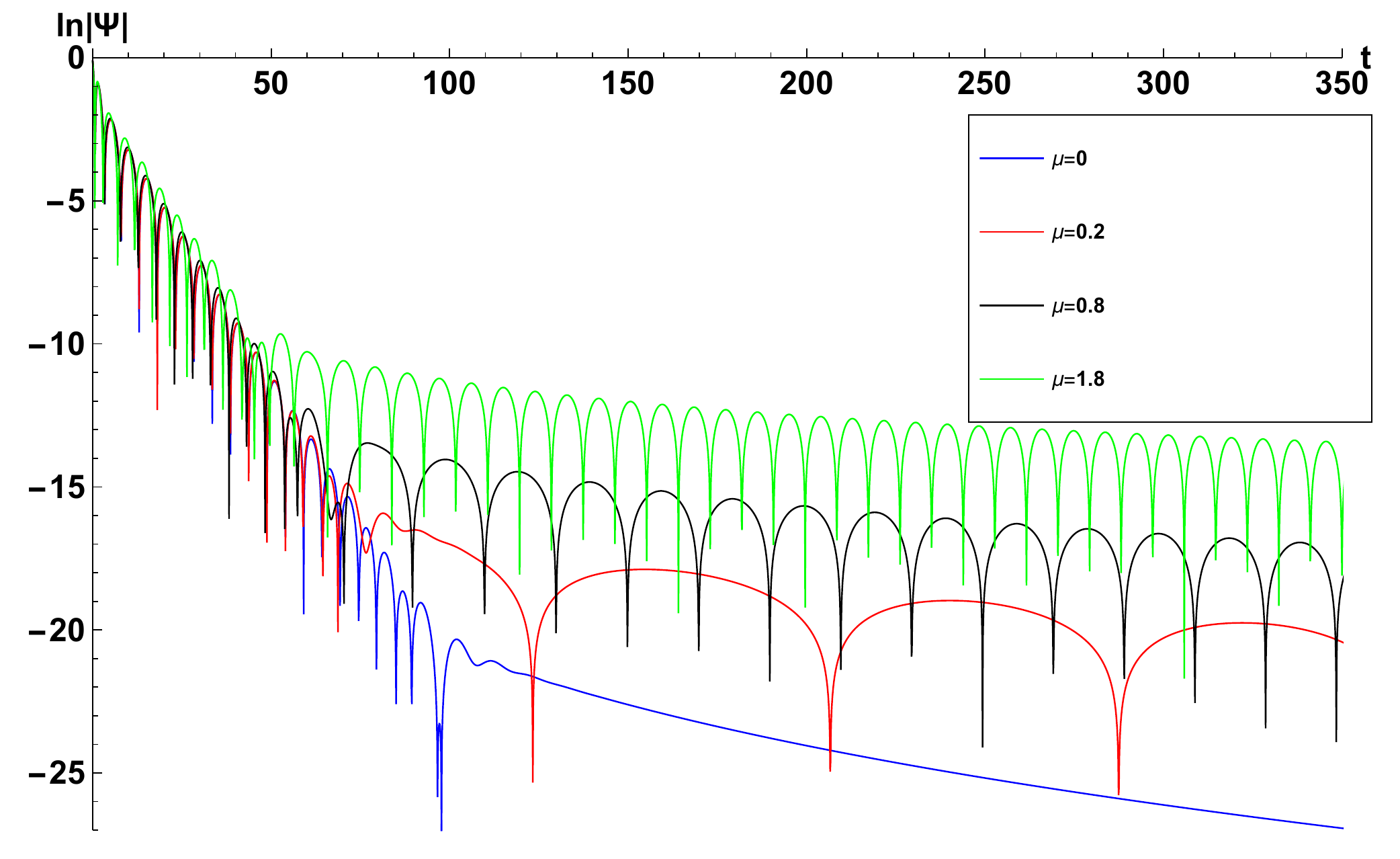}
\end{center}
\caption{The scalar field evolution with two families of oscillations represented: the photon sphere modes and the cosmological ones. The geometry parameters read $M=2Q=3.10^6 \Lambda = 1$ and the angular momentum of the field, $\ell = 1$.}
\label{fr3}
\end{figure}

Although the cosmological QNMs in RNdS geometry are dominant over the other two families for very small values of the cosmological constant, the photon sphere oscillations appear as the next sub-leading contribution. Thus, depending on the value of the mass and angular momentum of the field, we can see the photon sphere modes as the secondary part of the field evolution (after an initial burst), followed by the dominant cosmological frequencies as represented in Fig. \ref{fr3}. The fact that the ladder oscillates instead of exponentially decays is related to the presence of the mass of the scalar field: in the massless limit, we see a purely imaginary profile (see  Fig. \ref{fr3}).

In the small cosmological constant regime, the above-described modes dominate the field evolution. On the other way, for a near-extremal RNdS black hole, the purely imaginary frequencies similar to the one described in the previous section take control of the field profile. That is what we analyse in what follows.

\subsection{The near extremal modes}

Such family becomes representative when the event and Cauchy horizons approach, so that, the surface gravity of both structures is similar. The modes have the same expression as that for Reissner-Nordstr\"om black hole without $\Lambda$ as guidance \cite{hod2017quasibound, Fontana_2019} given by equation (\ref{rnf1}).

The qualitative behaviour does not bring a transition in the dominance of the QNMs as in \cite{Lagos_2020,Aragon:2020tvq}. Furthermore, this is the single family with only purely imaginary oscillations, for the massive scalar field in RNdS. According to (\ref{rnf1}), the higher the mass the higher the damping factor, for every $\ell$. This fact is observed in Fig. \ref{fr4}, for different $\ell$'s. The QNMs displayed in such figure are collected through the methods of integration in double null coordinates plus prony, as mentioned above, in the RNdS geometry, using the pure RN solution as guidance.
In the left panel of Fig. \ref{fr4}, we see that the purely imaginary modes increase with $\mu$. Those eigenvalues are observed in the final phase of the profiles as exponential decay, being the initial quasinormal ringing, the usual photon sphere oscillations (see right panel of Fig. \ref{fr4}).

Once we establish the regime at which the previous two families control the field evolution, outside of that, the field profile is dominated by the PS modes, studied in the next subsection.

\subsection{The photon sphere modes.}

This kind of modes are associated with the last unstable photon orbit around the black hole, and with the Lyapunov exponent of such orbits as described in \cite{Cardoso_2009}. They can be calculated with the WKB approximation with the improvements from the last years to 13th order \cite{Iyer:1986np,Iyer:1986nq,Kokkotas:1988fm,Seidel:1989bp,Konoplya_2003,Matyjasek:2017psv}.

\begin{figure*}
\begin{center}
\includegraphics[width=0.4\textwidth]{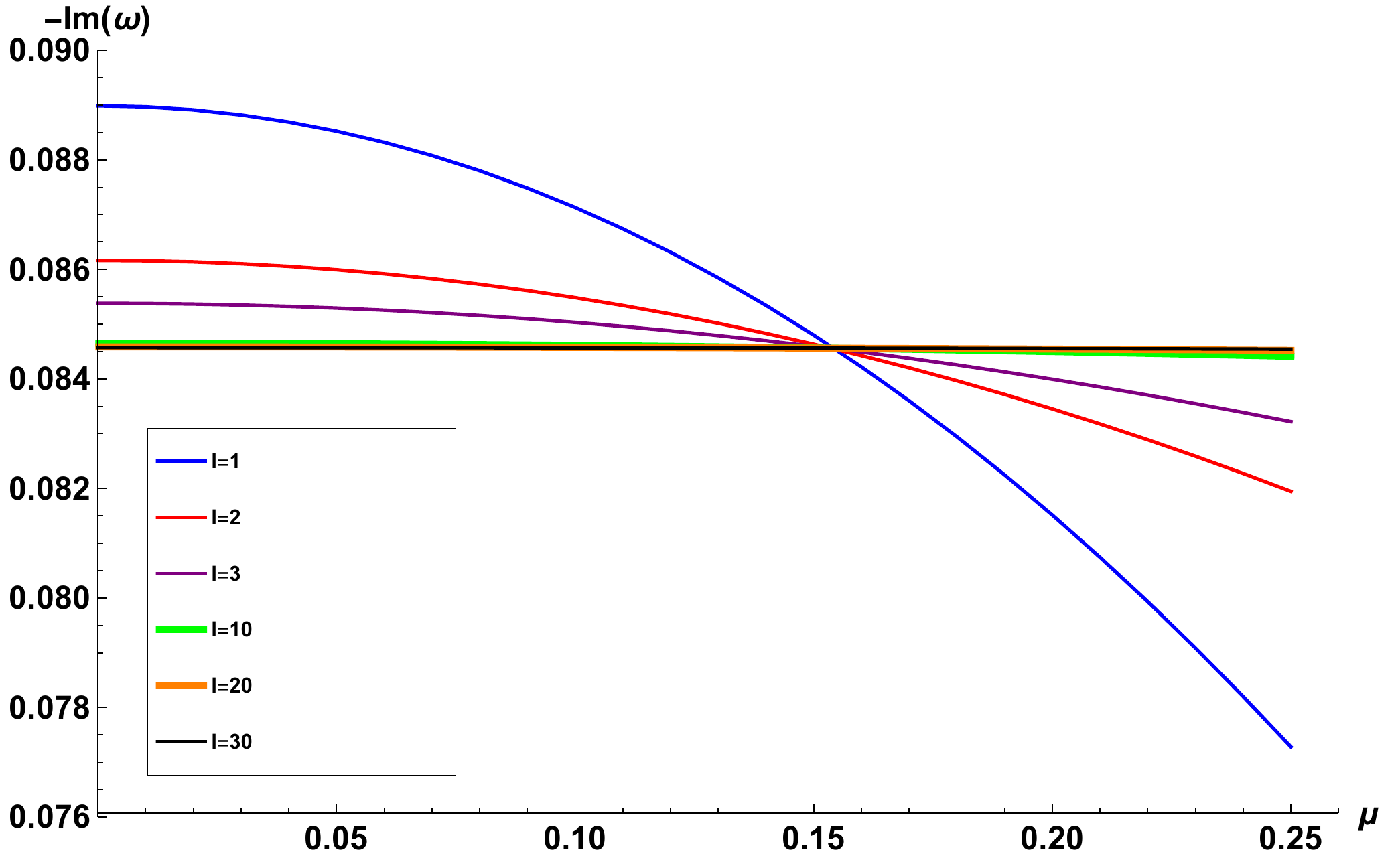}
\includegraphics[width=0.4\textwidth]{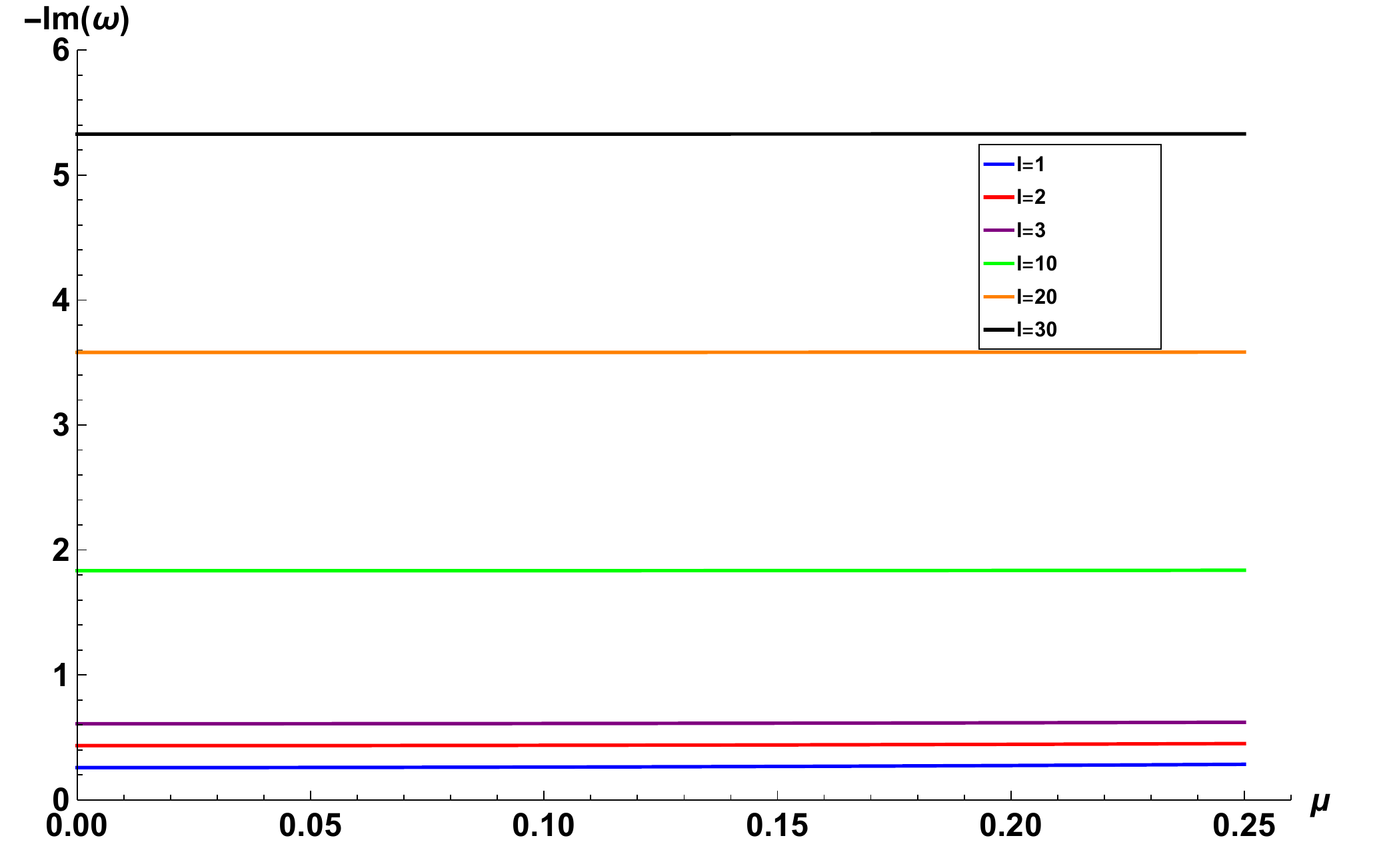}
\end{center}
\caption{The photon sphere modes for a massive scalar field in RNdS collected with the WKB approximation. The geometry parameters read $M = Q/2 = 100\Lambda /3 = 1$.}
\label{fr5}
\end{figure*}

Interestingly enough, we can see here that the anomalous interplay of dominance of the QNMs, is present, bringing along, the existence of a critical scalar field mass, what is represented in Fig. \ref{fr5}. The critical value $\mu \sim 0.16$ varies according to the geometric parameters. The real parts of the modes vary slightly with the change of $\mu$ (for small values): the scalar mass plays no role on the WKB QNMs. This is completely different from the cosmological modes behaviour, which scales $\mu$ to $Re (\omega )$.

Considering the descriptions of the above subsections, determine the lowest lying mode in RNdS is a trickier query, compared to RN. For one side, we point out the presence of the same kind of anomalous behaviour for the decay rate and the existence of a critical scalar field mass
reported for the Schwarzschild \cite{Lagos_2020} and Schwarzschild-de Sitter \cite{Aragon:2020tvq} black holes.

However, in RNdS when the field evolution is not controlled by photon sphere modes namely, whenever $\tilde{\delta} \equiv \frac{Q_{max}-Q}{Q_{max}} \gtrsim 10^{-2}$ and $\frac{\Lambda}{\Lambda_{max}} \gtrsim 10^{-2}$, we have a dominance of $\ell=0$ for the massless (and small masses) scalar field, and the eikonal modes control the higher mass regime of evolution. In an even more trickier way, for $\tilde{\delta} \lesssim 10^{-3}$, the NE modes take control over the evolution and $\ell =0$ is the frequency with the smallest $\omega_I$, whatever the mass of the scalar field.

The most complicated scenario is given by a small cosmological constant background, e. g.  $\frac{\Lambda}{\Lambda_{max}} \lesssim 10^{-3}$. In such case, the dominance depends on the scalar field mass (no interplay present as well). For small masses $\ell =0$ represents the lowest lying frequency. As from a specific value of $\mu$, $\ell=1$ controls the evolution. Then, approximately twice that mass, $\ell=2$ prevails. This behaviour seems to indicate that the eikonal modes are dominant for very high scalar field masses, and the smallest $\ell$'s overrule the small mass regime, though it is not possible to affirm the exact interval of dominance for each $\ell$.

Since the only group of frequencies in which the dominance turns over with the mass of the scalar field is the PS modes, in the next section we will analyse the WKB with its semi-analytic expansions.

\section{Analysis using the WKB method}
\label{WKBJ}

In order to get some analytical insight of the behaviour of the QNFs in the eikonal limit $\ell \rightarrow \infty$, we use the method based on Wentzel-Kramers-Brillouin (WKB) approximation initiated by Mashhoon \cite{1983mgm..conf..599M} and by Schutz and Iyer \cite{Schutz:1985zz}. Then, Iyer and Will computed the third order correction \cite{Iyer:1986np}, and Konoplya extended to the sixth order \cite{Konoplya:2003ii}, followed by its extension up to the 13th order \cite{Matyjasek:2017psv}, see also \cite{Konoplya:2019hlu}. This method has been used to determine the QNFs for asymptotically flat and asymptotically de Sitter black holes. This is due to the fact that the WKB method can be used for effective potentials which have the form of a barrier potential, approaching to a constant value at the event horizon and spatial infinity \cite{Konoplya:2011qq}.
The QNMs are determined by the behaviour of the effective potential near its maximum value $r^*_{max}$. The Taylor series expansion of the potential around its maximum is given by
\begin{equation}
V(r^*)= V(r^*_{max})+ \sum_{i=2}^{i=\infty} \frac{V^{(i)}}{i!} (r^*-r^*_{max})^{i}   \,,
\end{equation}
where
\begin{equation}
V^{(i)}= \frac{d^{i}}{d r^{*i}}V(r^*)|_{r^*=r^*_{max}}
\end{equation}
corresponds to the $i$-th derivative of the potential with respect to $r^*$ evaluated at the maximum of the potential $r^*_{max}$. Using the WKB approximation up to 6th order the QNFs are given by the following expression \cite{Hatsuda:2019eoj}

\begin{equation} \label{omega}
\omega^2 = V(r^*_{max}) -2 i U \,,
\end{equation}
where
\begin{multline}
 U = N\sqrt{-V^{(2)}/2}\\
+\frac{i}{64} \left( -\frac{1}{9}\frac{V^{(3)2}}{V^{(2)2}} (7+60N^2)+\frac{V^{(4)}}{V^{(2)}}(1+4 N^2) \right)  \\
+\frac{N}{2^{3/2} 288} \Bigg( \frac{5}{24} \frac{V^{(3)4}}{(-V^{(2)})^{9/2}} (77+188N^2) \\
+\frac{3}{4} \frac{V^{(3)2} V^{(4)}}{(-V^{(2)})^{7/2}}(51+100N^2)  + \frac{1}{8} \frac{V^{(4)2}}{(-V^{(2)})^{5/2}}(67+68 N^2)\\
+\frac{V^{(3)}V^{(5)}}{(-V^{(2)})^{5/2}}(19+28N^2)+\frac{V^{(6)}}{(-V^{(2)})^{3/2}} (5+4N^2)  \Bigg)\,,
\end{multline}
and $N=n_{PS}+1/2$, with $n_{PS}=0,1,2,\dots$, the overtone number.
\newline

Defining $L^2= \ell (\ell+1)$, we find that for large values of $L$ and low values of $Q$, the maximum of the potential is approximately at
\begin{equation}
\notag r_{max} \approx r_0+ \frac{1}{L^2} r_1 + \mathcal{O}(L^{-4})~,
\end{equation}
where
\begin{multline}
 r_0 \approx  \frac{1}{2} \left( 3M+\sqrt{9M^2-8Q^2} \right) \approx 3M-\frac{2Q^2}{3M} - \frac{4 Q^4}{27 M^3} + \mathcal{O}(Q^5)\,, \\
 r_1 \approx -\frac{M}{3} \left(1-27 \mu^2 M^2+18 \Lambda M^2 \right) \left(1- 9 \Lambda M^2 \right) \\
+\frac{\left(5-72 \Lambda M^2\right) \left(1 -27 \mu^2 M^2 +18 \Lambda M^2 \right)}{27 M}Q^2 \\
 + \left(  4 M \Lambda \left( -3 \mu^2 +2 \Lambda \right) +  \frac{4}{81 M^3}\right) Q^4 +\mathcal{O}(Q^5)\,,
\end{multline}
and
\begin{equation} \label{coa}
\notag V(r^*_{max}) \approx V_0 L^2+ V_1 +\mathcal{O}(L^{-2}) \,,
\end{equation}
where
\begin{eqnarray}
\nonumber
 V_0 \approx  \frac{1}{27 M^2} \left(1-9 \Lambda M^2 \right) +\frac{Q^2}{81 M^4} +\frac{4 Q^4}{729 M^6} + \mathcal{O} (Q^5)~, \\
\nonumber
 V_1 \approx  \frac{\left(2+27\mu^2 M^2-18 \Lambda M^2 \right) \left(1-9 \Lambda M^2 \right)}{81 M^2} \\
\nonumber
+ \frac{4+36 \Lambda M^2-648 \Lambda^2 M^4-27 \mu^2M^2 \left(1-36 \Lambda M^2 \right)}{729 M^4} Q^2  \\
\nonumber
+  \frac{2 \left( 5 +72\Lambda M^2 -324 \Lambda^2 M^4 -54 \mu^2 M^2 \left( 1- 9 \Lambda M^2 \right) \right)}{6561 M^6} Q^4 \\ \nonumber + \mathcal{O} (Q^5)~,\\
\end{eqnarray}

while the second derivative of the potential evaluated at $r^*_{max}$ is given by
\begin{equation}
V^{(2)}(r^*_{max}) \approx V^{(2)}_0 L^2 +V^{(2)}_1 + \mathcal{O}(L^{-2}) \,,
\end{equation}
where
\begin{eqnarray}
\nonumber
 V^{(2)}_0   \approx  -\frac{2 \left(  1- 9 \Lambda M^2\right)^2}{729 M^4} - \frac{4 \left( 1- 9\Lambda M^2 \right) \left( 2 +9\Lambda M^2 \right)}{6561 M^6} Q^2 \\
 \nonumber - \frac{2 \left( 13+36\Lambda M^2 -648 \Lambda^2 M^4 \right)}{59049 M^8} Q^4 +\mathcal{O} (Q^5)~,   \\
\\
\nonumber
V^{(2)}_1  \approx  -\frac{(16M^{-2}+36 \Lambda  -1620 \Lambda^2 M^2 -54 \mu^2  (2-45 \Lambda M^2) )}{6561 M^2( 1- 9 \Lambda M^2 )^{-2} } \\
\nonumber
- \frac{2 (1- 9 \Lambda M^2)}{59049 M^6} \Big(  22+ 270 \Lambda M^2 + 2916 \Lambda ^2 M^4-64152 \Lambda^3 M^6  \\
\nonumber
 -27 \mu^2 M^2 (1+225 \Lambda M^2 -3564 \Lambda^2 M^4) \Big) Q^2 \\
\nonumber
- \frac{2}{177147 M^8} \Big(2 (5-9\Lambda M^2  -324 \Lambda^2 M^4) (1+18 \Lambda M^2)^2 \\
\nonumber
+ 27 \mu^2 M^2 ( 4 - 279 \Lambda M^2 +1620 \Lambda^2 M^4 +11664 \Lambda^3 M^6)   \Big) Q^4 \\+ \mathcal{O} (Q^5)~,\hspace{2.0cm}
\end{eqnarray}
and the higher derivatives of the potential evaluated at $r^*_{max}$ yields the following expressions
\begin{align}
\nonumber
 V^{(3)}(r^*_{max}) \approx  (1-9\Lambda M^2)^3 \Bigg(\frac{4 }{6561M^5}+ \frac{4 (1+18\Lambda M^2)}{19683 M^7} Q^2  \\
\nonumber
-\frac{4 (1-135 \Lambda M^2+1458 \Lambda^2 M^4 -2916 \Lambda^3 M^6)}{177147 M^9(1-9\Lambda M^2)^{-3}}Q^4  \\
 +\mathcal{O}(Q^5)  \Bigg) L^2+ \mathcal{O}(L^0)\,, \hspace{2.0cm} \\
\nonumber
 V^{(4)}(r^*_{max})  \approx  \Bigg( \frac{16 (1-9 \Lambda M^2 )^3}{19683M^6} \\
\nonumber
+\frac{8 (1-9 \Lambda M^2)^2 (11+54 \Lambda M^2 +81 \Lambda^2 M^4)}{177147 M^8} Q^2  \\
\nonumber
+\frac{16 (7 -27 \Lambda M^2-486 \Lambda^2 M^4+1458 \Lambda^3 M^6)}{531441 M^{10}}  Q^4 \\
+\mathcal{O}(Q^5)  \Bigg)   L^2+ \mathcal{O}(L^0)\,, \hspace{2.0cm} \\
\nonumber
 V^{(5)}(r^*_{max}) \approx  \Bigg(-\frac{40 (1-9 \Lambda M^2)^4}{59049 M^7} \\
\nonumber
-\frac{160 (1- 9\Lambda M^2)^3 (1+18 \Lambda M^2)}{531441 M^9} Q^2 \\
\nonumber
+ \frac{80 (1-9 \Lambda M^2)^2 (1-288 \Lambda M^2 +324 \Lambda^2 M^4)}{4782969 M^{11}} Q^4     \\
+ \mathcal{O} (Q^5) \Bigg) L^2+ \mathcal{O}(L^0)\,,   \hspace{2.0cm} \\
\nonumber
  V^{(6)}(r^*_{max}) \approx  \Bigg(  -\frac{16(1-9\Lambda M^2)^4(4+15 \Lambda M^2)}{177147 M^8} \\
\nonumber
-\frac{16 (1-9\Lambda M^2)^3(113+45 \Lambda M^2+6966 \Lambda^2 M^4)}{4782969 M^{10}} Q^2 \\
\nonumber
- \frac{32  (367-1701\Lambda M^2+18225\Lambda^2 M^4 +40824 \Lambda^3 M^6)}{43046721 M^{12}(1-9 \Lambda M^2)^{-2}} Q^4\\
+ \mathcal{O}(Q^5) \Bigg) L^2 + \mathcal{O}(L^0)\,.\hspace{2.0cm}
\end{align}

Using these results together with Eq. (\ref{omega}) we obtain the following analytical QNFs valid for small values of $Q/M$ and large values of $L$
\begin{equation}
\omega \approx \omega_1 L+ \omega_0 +\omega_{-1} L^{-1} + \omega_{-2} L^{-2} + \mathcal{O} (L^{-3})\,,
\end{equation}
where
\begin{eqnarray}
\nonumber
 \omega_1  \approx  \frac{\sqrt{1-9 \Lambda M^2}}{3 \sqrt{3} M}+ \frac{Q^2}{18 M^3 \sqrt{3(1-9 \Lambda M^2)}} \\
\nonumber
+ \frac{(13-144 \Lambda M^2)Q^4}{648 \sqrt{3}M^5 (1- 9 \Lambda M^2)^{3/2}} + \mathcal{O}(Q^5)\,,  \\
\nonumber
 \omega_0 \approx  -i \frac{\sqrt{1-9 \Lambda M^2}}{6 \sqrt{3}M}-i \frac{(1+18 \Lambda M^2)Q^2}{108M^3\sqrt{3(1-9 \Lambda M^2)}} \\
\nonumber
+i \frac{(1-18 \Lambda M^2)^2 Q^4}{432 M^5 \sqrt{3(1-9 \Lambda M^2)}(1- 9 \Lambda M^2)}+ \mathcal{O}(Q^5)\,, \\
\nonumber
 \omega_{-1}  \approx  \frac{(34 + 9M^2 (108 \mu^2 -61 \Lambda)) \sqrt{ 1-9 \Lambda M^2}}{648 M \sqrt{3}}+\\
\nonumber
 \frac{(122+  9M^2 (\Lambda (251-4086 \Lambda M^2) -108 \mu^2 (5-72 \Lambda M^2)))Q^2}{11664 M^3 \sqrt{3 (1-9\Lambda M^2 )}}  \\
\nonumber
+ \frac{ (1550-27 M^2 (\Lambda (-139+36 \Lambda M^2 (313-1524 \Lambda M^2)))) Q^4}{419904 M^5 \sqrt{3 (1- 9\Lambda M^2)} (1-9 \Lambda M^2)} \\
\nonumber
- \frac{ (27 M^2 (-36 \mu^2 (41+288 \Lambda M^2 (-2+9 \Lambda M^2)))) Q^4}{419904 M^5 \sqrt{3 (1- 9\Lambda M^2)} (1-9 \Lambda M^2)} \\
\nonumber
+ \mathcal{O}(Q^5)\,, \hspace{2.0cm}\\
\nonumber
 \omega_{-2}  \approx -i \frac{(137 -45M^2 (648\mu^2-401 \Lambda)) (1- 9 \Lambda M^2 )^{3/2}}{ 23328 \sqrt{3}M} \\
\nonumber
+i \frac{(145 + \Lambda(-1265 +31098 \Lambda M^2)))(1- 9\Lambda M^2)^{1/2} Q^2}{139968 \sqrt{3} M^3} \\
\nonumber
-i \frac{(27M^2 (216 \mu^2 (11-234 \Lambda M^2)))(1- 9\Lambda M^2)^{1/2} Q^2}{139968 \sqrt{3} M^3} \\
\nonumber
 +i\frac{(1505 + 3M^2 (5832 \mu^2 (1+6 \Lambda M^2)(-11+ 126 \Lambda M^2))) Q^4}{1679616 M^5 \sqrt{3(1- 9\Lambda M^2)}} \\
\nonumber
 +i\frac{( 3M^2 (-7\Lambda (-4469+36 \Lambda M^2 (790+10593 \Lambda M^2)))) Q^4}{1679616 M^5 \sqrt{3(1- 9\Lambda M^2)}}\\+ \mathcal{O}(Q^5)\,. \hspace{2.0cm}
\end{eqnarray}

\begin{figure*}
\begin{center}
\includegraphics[width=0.48\textwidth]{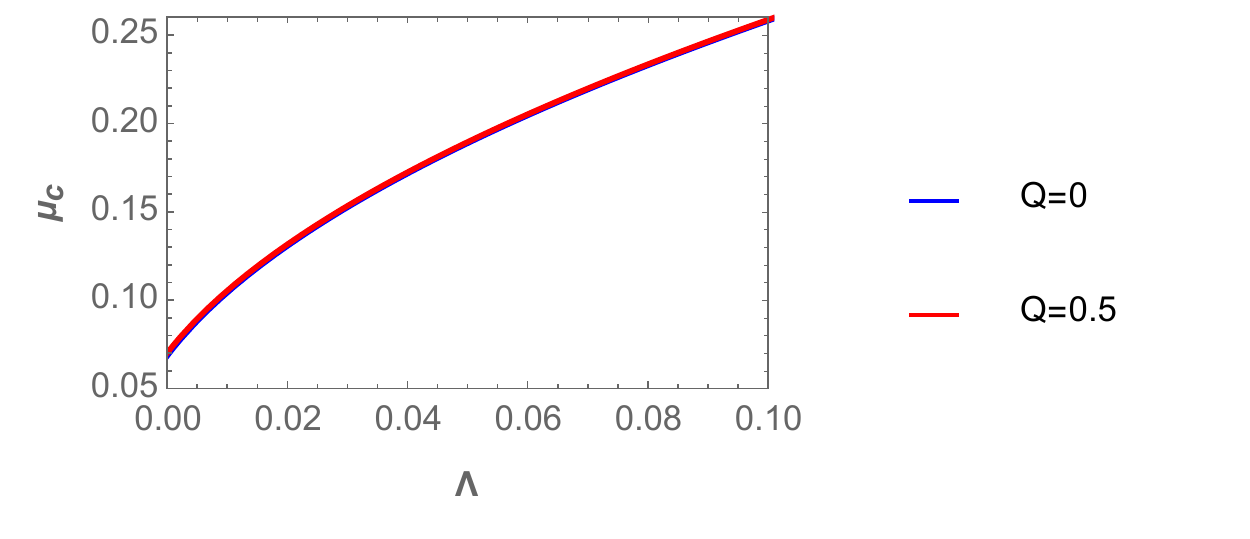}
\includegraphics[width=0.48\textwidth]{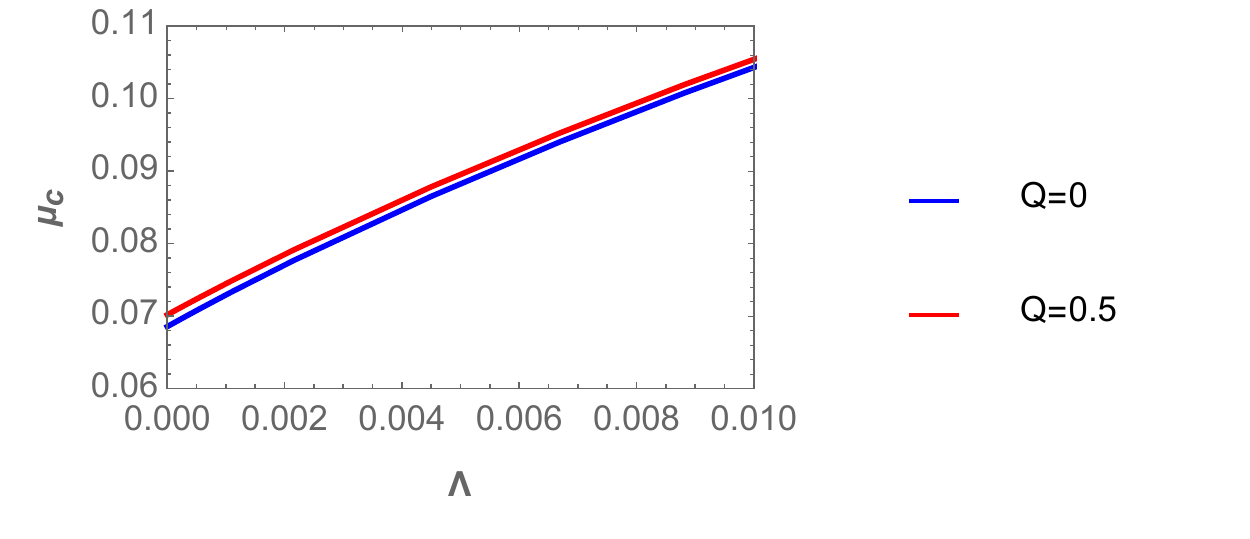}
\end{center}
\caption{The behaviour of the critical scalar field mass as a function of the cosmological constant, with $M=1$. Left figure for a global behaviour and right figure for a zone near to null cosmological constant.}
\label{WKB}
\end{figure*}

The term proportional to $1/L^2$ is zero at the value of the critical mass $\mu_c$, which is given by
\begin{eqnarray}
\label{mc}
\nonumber
\mu_c M  \approx \frac{\sqrt{137+18045 \Lambda M^2 }}{54 \sqrt{10}} \\
\nonumber
+ \frac{391 + 1350 \Lambda M^2}{1620 \sqrt{10} \sqrt{137+18045\Lambda M^2} M^2}Q^2 - \\
\nonumber
  \frac{3277- 24526740 \Lambda M^2 -772308000 \Lambda ^2 M^4}{97200 \sqrt{10} (137+18045 \Lambda M^2)^{3/2} M^4} Q^4 + \mathcal{O}(Q^5)\,, \\
\end{eqnarray}
and it is valid for small values of $Q/M$ and $n_{PS}=0$ {\bf{\footnote{For Schwarzschild dS black holes \cite{Aragon:2020tvq}, it was shown that for small values of $M^2\Lambda$ the critical mass depends on the overtone number $n_{PS}$, that is, the critical mass value increases when the overtone number increases. However, when $M^2\Lambda$ increases, the black hole becomes extremal when $9 M
^2 \Lambda=1$, and the critical mass value does not depend on the overtone number. The same behaviour was observed in the context of $f(R)$ gravity \cite{Aragon:2020xtm}.}}}. For $Q=0$ we recover the result of critical mass for Schwarzschild-dS black holes \cite{Aragon:2020tvq}. In Fig. \ref{WKB} we show the behaviour of the critical scalar field mass given by Eq. (\ref{mc}) as a function of the cosmological constant, we can observe that the value of the critical scalar field mass increases when the cosmological constant increases, and when the charge of the black hole increases.

The results displayed in Fig. \ref{WKB} left panel, with the above semi-analytical expansions of the WKB approximation substantiate those in the previous section. In Fig. \ref{fr5} left panel, we see $\mu_c \sim 0.16$ for $Q\sim 0.5$ - also true for smaller black hole charges -  corresponding approximately to the same point of \ref{WKB} when $\Lambda = 0.03$.

The last spacetime we analyse is the charged geometry with negative cosmological constant, in which the WKB approximation is not valid for $V\rightarrow \infty$ in $\mathcal{I}$.

\begin{figure*}
\begin{center}
\includegraphics[width=0.47\textwidth]{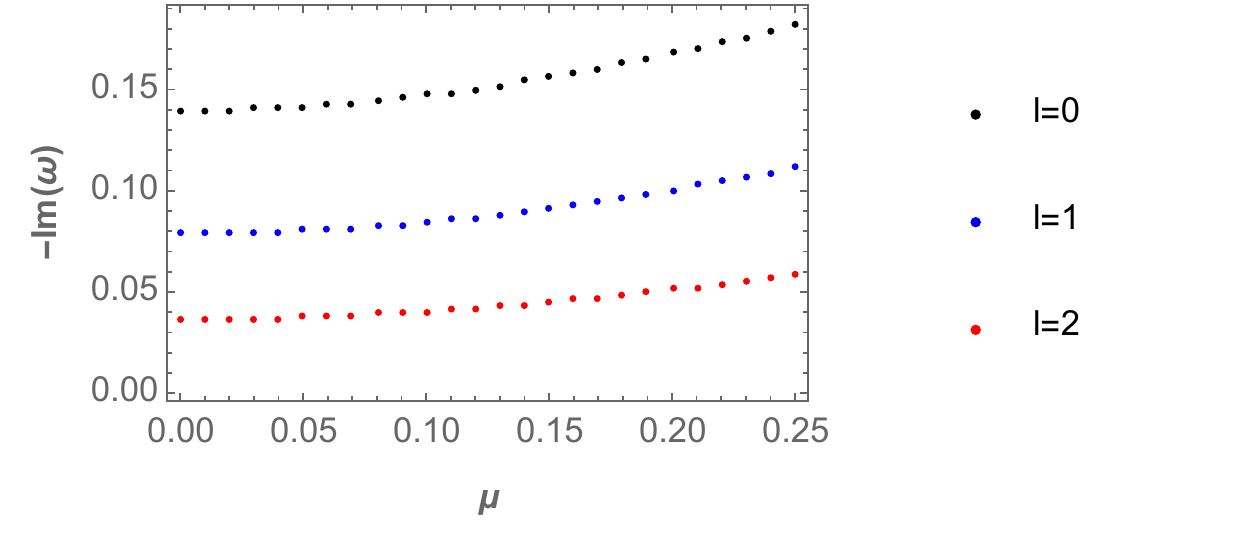}
\includegraphics[width=0.47\textwidth]{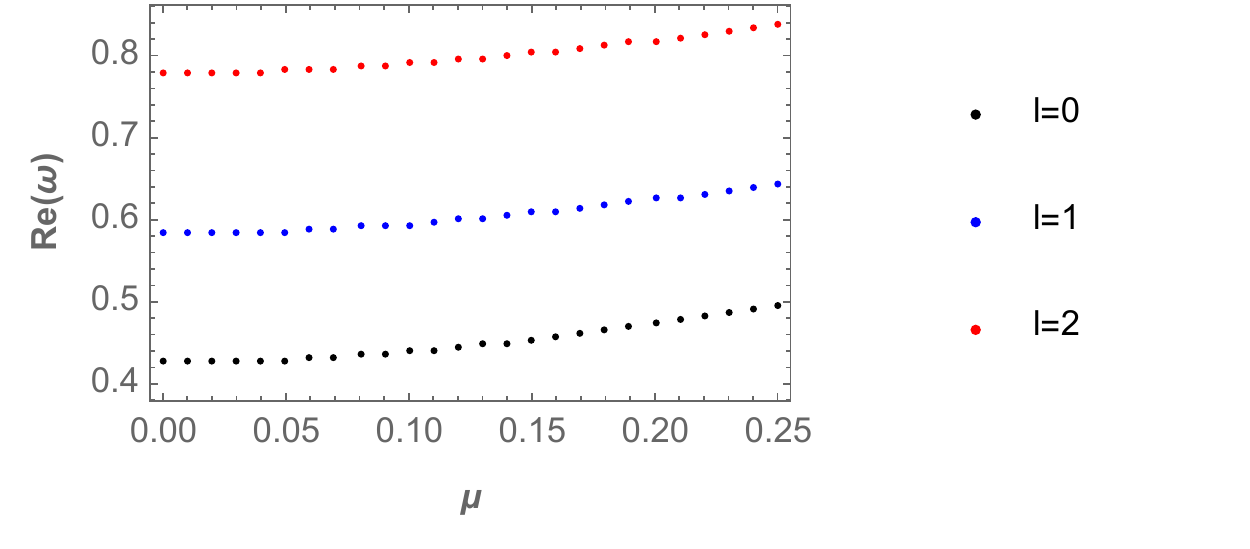}
\end{center}
\caption{The behaviour of the fundamental mode $-Im(\omega)$ (left figure)  and $Re(\omega)$ (right figure) as a function of $\mu$ for different values of the parameter $\ell=0,1,2$, with $M=1$, $Q=1/2$, and $\Lambda=-1/10$.  $r_-\approx 0.134$ , and $r_+ \approx 1.691$.}
\label{F1B1}
\end{figure*}

\begin{figure*}
\begin{center}
\includegraphics[width=0.47\textwidth]{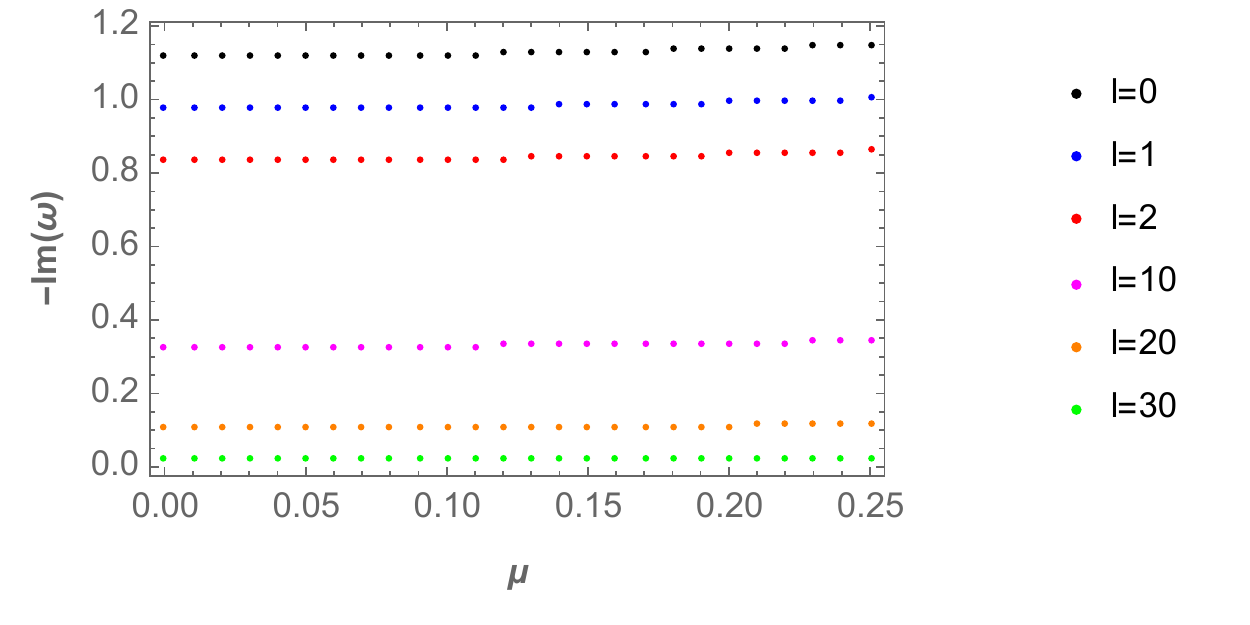}
\includegraphics[width=0.47\textwidth]{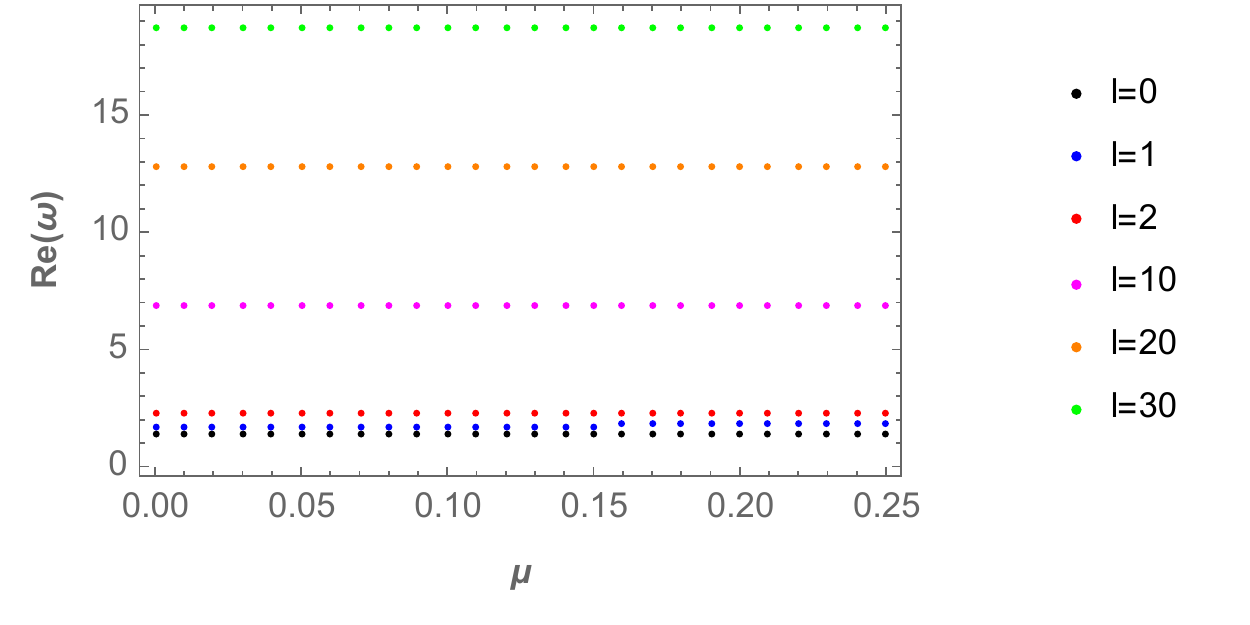}
\end{center}
\caption{The behaviour of the fundamental mode $-Im(\omega)$ (left figure)  and $Re(\omega)$ (right figure) as a function of $\mu$ for different values of the parameter $\ell=0,1,2,10,20,30$, with $M=1$, $Q=1/2$, and $\Lambda=-1$. $r_-\approx 0.134$ , and $r_+\approx 1.207$.}
\label{F1B11}
\end{figure*}

\section{Reissner-Nordstr\"om Anti de Sitter quasinormal modes}
\label{RNAdScase}

In this section, we use the pseudospectral Chebyshev method \cite{boyd01}
that represents an effective method to find high overtone modes in RNAdS geometry \cite{Finazzo:2016psx,Gonzalez:2017shu,Gonzalez:2018xrq,Becar:2019hwk,Aragon:2020qdc}. Under the change of variable $y=1-r_H/r$ the radial wave equation (\ref{radial}) becomes
\begin{eqnarray}
\nonumber
(1-y)^4 f(y) R''(y) +(1-y)^4 f'(y)R'(y) \\
\label{r}
+ \left( \frac{\omega^2 r_H^2}{f(y)}-
\ell(\ell+1) (1-y)^2 -\mu^2 r_H^2 \right) R(y)=0\,,
\end{eqnarray}
where the prime denotes derivative with respect to $y$, the event horizon is located at $y=0$ and the spatial infinity at $y=1$. 
By considering the behaviour of the field at the event horizon (only ingoing waves) and at infinity (vanishing scalar field), it is possible to define the following ansatz
\begin{equation}
R(y)= e^{-\frac{i \omega r_H}{f'(0)} \ln{y}} (1-y)^{\frac{3}{2} + \sqrt{\left(\frac{3}{2}  \right)^2 -\frac{3 \mu^2}{\Lambda}}} F(y) \,.
\end{equation}

By inserting the above ansatz for $R(y)$ in Eq. (\ref{r}), a differential equation for the function $F(y)$ is obtained. The solution for the function $F(y)$ is assumed to be a finite linear combination of the Chebyshev polynomials and can be plugged into the differential equation for $F(y)$. Also, the interval $[0,1]$ is discretized at the Chebyshev collocation points. After that, the differential equation is evaluated at each collocation point, yielding a system of algebraic equations, for a generalized eigenvalue problem, that can be solved numerically to obtain the QNFs.

Now, in order to see if there is an anomalous decay rate of quasinormal modes, we plot in Fig. \ref{F1B1}, the behaviour of the fundamental QNFs, for small values of the parameter $\ell$, and different values of $\mu$.
We can observe an anomalous behaviour of the QNMs in Reissner-Nordstr\"om-AdS black holes because the longest-lived modes are always the ones with higher angular number; however, there is not a critical mass where the behaviour of the modes is inverted, thus no interplay of modes relative to $\ell$. The same behaviour is observed, when we increase the absolute value of the cosmological constant, and higher-values of the angular number. Also, the decay rate and the frequency of the oscillations increases when $\Lambda$ increases, see  Fig. \ref{F1B11}. On the other hand, when the black hole charge increases, see Table \ref{AdS}, $\Delta (r)=r_+-r_-$ decreases (near extremal modes), and for small values of the angular number the dominant mode becomes purely imaginary (just damping), while that for higher-values of the angular number the dominant modes present oscillatory QNMs.

\begin {table*}
\caption {Fundamental modes ($n =
      0 $) for massless scalar fields with $\ell = 0$ and $10$,  in the background of a RNAdS black hole with $M=1$, $\Lambda = -1 $, and different values of the black hole charge $Q$.}
\label {AdS}\centering
\begin {tabular} { | c | c | c | c |}
\hline
${}$ & $Q = 0.50 $ & $Q= 0.70 $ & $Q= 0.80 $ \\\hline
$r_-$ &
$0.134$ &
$0.287$ &
$0.408$  \\\hline
$r_+$ &
$1.207$ &
$1.106$ &
$1.019$  \\\hline
$\omega(\ell=0)$ &
$1.37902718 - 1.11707063 i$ &
$1.26914155 - 1.11549524 i$ &
$1.20019018 - 1.15709954 i$  \\\hline
$\omega(\ell=10)$ &
$6.87520834 - 0.32749992 i$ &
$6.86928091 - 0.28327488 i$ &
$6.86569597 - 0.25173534 i$  \\\hline
${}$ & $Q= 0.85 $ & $Q= 0.90 $ & $Q= 0.91 $ \\\hline
$r_-$ &
$0.492$ &
$0.629$ &
$0.689$ \\\hline
$r_+$ &
$0.953$ &
$0.837$ &
$0.781$ \\\hline
$\omega(\ell=0)$ &
$-1.146660864 i$ &
$-0.425213719 i$ &
$-0.180659019 i$ \\\hline
$\omega(\ell=10)$ &
$6.86375299 - 0.23266374 i$ &
$6.86167698 - 0.21072458 i$ &
$6.86123870 - 0.20592740 i$  \\\hline
\end {tabular}
\end{table*}

In table \ref{AdS} we clearly see the existence of two quasinormal behaviours, where the field profile dominance is controlled by an oscillatory or a purely imaginary $\omega$. The change in dominance happens at high charges, strongly indicating the presence of a near extremal family of modes also in RNAdS geometry. This family is confirmed to be present in the very near-extremal regime: taking as an example $M=-\Lambda=1$, $\mu = \ell = 0$ and $\tilde{\delta}=10^{-5}$, we obtain for the lowest lying QNM with the above numerics $\omega = -0.003791i$, very near to the result of relation (\ref{rnf1}) given by $\omega = -0.003781i$.

\begin{figure*}
\begin{center}
\includegraphics[width=0.47\textwidth]{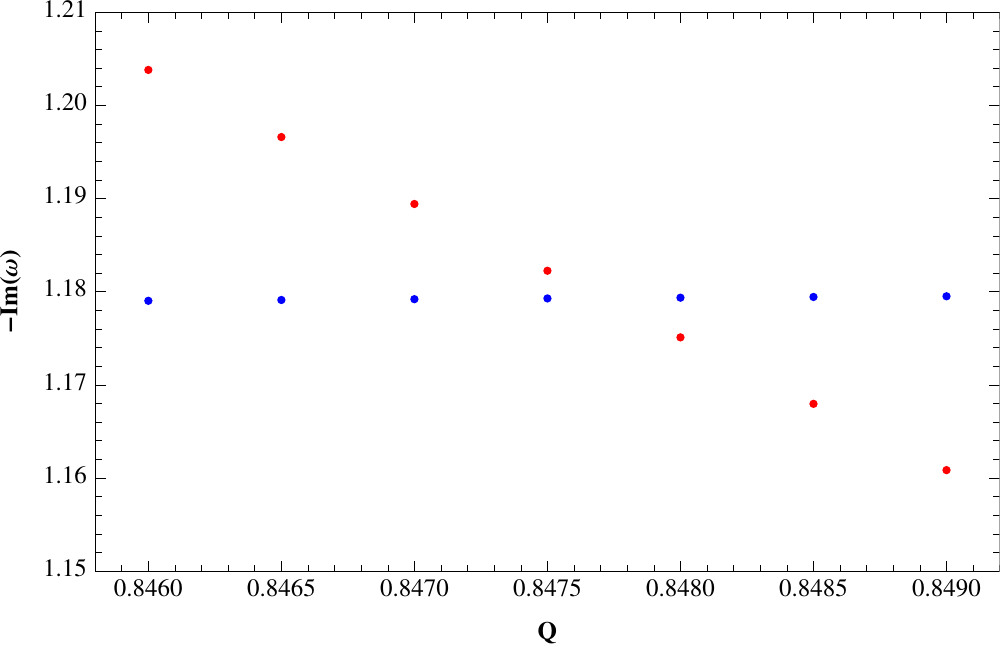}
\end{center}
\caption{The behaviour of $-Im(\omega)$ as a function of $Q$ for different values of the overtone number $n=0,1$, with $M=1$, $\ell=0$, and $\Lambda=-1$. Red points for purely imaginary modes and blue points for oscillatory modes.}
\label{Qcritical}
\end{figure*}

Now, in order to determine the critical value of the charge $Q_c$, i.e, the value of $Q$ for which the field profile dominance change from oscillatory to purely
imaginary modes (fixed $\ell=0$), we display its behaviour in Fig. \ref{Qcritical}, according to which $Q_c \approx 0.848$ for zero angular momentum. Thus, whenever $Q<Q_c$ the field profile is controlled by oscillatory modes, while higher values of charge, $Q\geq Q_c$, yields a field profile dominated by purely imaginary modes. 
Table \ref{AdS} shows that for high values of $\ell$, the field profile is controlled by oscillatory modes. By lowering the angular momentum (fixed $Q$), we can always find a critical value $\ell_c$ where a transition from purely imaginary to oscillatory modes takes place (as long as $Q\geq Q_c$). Thus, for $Q=0.85$, we have
$\ell_c=0$, for $Q=0.90$, $\ell_c=1$, and for $Q=0.9115$, $\ell_c=4$. Hence $\ell_c$ increases with the increment of the charge of the black hole, which can be seen in Table \ref{AdS}.

\begin {table*}
\caption {Fundamental modes ($n =
      0 $) for massless scalar fields with $Q = 0.80$, $0.85$, $0.90$, and $0.9115$,  in the background of a RNAdS black hole with $M=1$, $\Lambda = -1 $, and different values of $\ell$.}
\label {AdS}\centering
\begin {tabular} { | c | c | c |}
\hline
${}$ & $Q = 0.80$ & $Q=0.85$   \\\hline
$\omega(\ell=0)$ &
$1.20019018 - 1.15709954 i$ &
$-1.146660864 i$  \\\hline
$\omega(\ell=1)$ &
$1.60914643 - 0.92104166 i$ &
$1.57255034 - 0.91943475 i$   \\\hline
$\omega(\ell=2)$ &
$2.15350014 - 0.75609582 i$ &
$2.12445414 - 0.74098895 i$  \\\hline
$\omega(\ell=3)$ &
$2.72797216 - 0.63957929 i$ &
$2.70597899 - 0.62051031 i$  \\\hline
$\omega(\ell=4)$ &
$3.31256718 - 0.55052304 i$ &
$3.29564746 - 0.53016647 i$  \\\hline
$\omega(\ell=5)$ &
$3.90156873 - 0.47886639 i$ &
$3.88845585 - 0.45816008 i$  \\\hline
$\omega(\ell=10)$ &
$6.86569597 - 0.25173534 i$ &
$6.86375299 - 0.23266374 i$   \\\hline
$\omega(\ell=20)$ &
$12.79922150 - 0.05719447 i$ &
$12.80755770 - 0.04593707 i$   \\\hline
$\omega(\ell=30)$ &
$18.6958490 - 0.0049672 i$ &
$18.7039617 - 0.0027929 i$  \\\hline
${}$ & $Q = 0.90$ & $Q=0.9115$  \\\hline
$\omega(\ell=0)$ &
$-0.425213719 i$ &
$-0.110150970 i  $ \\\hline
$\omega(\ell=1)$ &
$-0.672442143 i$ &
$-0.17475212 i  $ \\\hline
$\omega(\ell=2)$ &
$2.09080984 - 0.72743641 i$  &
$ -0.24789205 i $ \\\hline
$\omega(\ell=3)$ &
$2.67958715 - 0.60067352 i$  &
$-0.32241548 i$ \\\hline
$\omega(\ell=4)$ &
$3.27514490 - 0.50789584 i$  &
$-0.39743177 i$ \\\hline
$\omega(\ell=5)$ &
$3.87255207 - 0.43496155 i$  &
$3.868393927 -0.429263545 i$ \\\hline
$\omega(\ell=10)$ &
$6.86167698 - 0.21072458 i$  &
$6.86117209-0.20519472 i$ \\\hline
$\omega(\ell=20)$ &
$12.81779480 - 0.03401161 i$  &
$12.82047962-0.03120091 i$ \\\hline
$\omega(\ell=30)$ &
$18.7124077 - 0.0012442 i$  &
$18.71437061-0.00098413 i$ \\\hline
\end {tabular}
\end{table*}

\clearpage

\section{Conclusions}
\label{conclusion}

In this work, we considered the Reissner-Nordstr\"om, the Reissner-Nordstr\"om-dS and the Reissner-Nordstr\"om-AdS black holes as background studying the propagation of massive scalar fields and its QNFs. The analyses we performed were threefolded: we investigate the existence of an anomalous decay rate behaviour, the presence of a critical scalar field mass, as well as, the dominant mode/family for massless and massive scalar fields in all three geometries.

For the Reissner-Nordstr\"om and Reissner-Nordstr\"om-dS solutions, more than one family of modes is present, and their behaviour by increasing the scalar field mass is different for each family. For the photon sphere modes, in both spacetimes, we locate an interplay of dominance in the spectra from high to small $\ell$ as we go from small to high scalar field masses. For the values of mass, charge, cosmological constant, and angular number considered here, the critical scalar field mass for the RN black hole has been obtained, increasing with the raise
 of the black hole charge. However, in the very near-extremal regime (e. g. $\tilde{\delta} < 10^{-3}$),
 the secondary family takes control over the field evolution and no interplay takes place,
 thus not presenting an anomalous behaviour in such limit.

The same feature happens in the very near-extremal RNdS black hole: the dominance is taken by the near-extremal family of modes, preventing the presence of such anomaly. In the photon sphere modes, otherwise, we expect higher critical masses than that of the RN metric. A third family of oscillations is existent when $\Lambda >0$, reported to dominate the field profiles for very small $\Lambda$'s. In such cases also no anomalous behaviour of the decay rate is present, although the behaviour of the modes with increasing $\mu$ is trickier than that of the near-extremal family. In that case, for different masses, different angular eigenfunctions dominates the profile, presenting the lowest $\omega_I$.

The understanding of such results was deepened as we applied semi-analytical expansions through the WKB analysis in order to get some analytical insight about the critical scalar field mass for Reissner-Nordstr\"om, and Reissner-Nordstr\"om-dS geometries. We have obtained the critical scalar field mass for small values of $Q/M$, and we have shown that the effect of the charge is to increase the value of the critical mass. Also, when the cosmological constant increases the critical scalar field mass increases.

In the case of a  Reissner-Nordstr\"om-AdS black hole background we have shown that decay rate of QNMs  of massive scalar fields under Dirichlet boundary conditions present an anomalous behaviour because the longest-lived modes are always the ones with higher angular number. Also, the decay rate  increases when the scalar field mass increases. However,  we showed that for large and intermediate black holes there is no critical scalar field mass where the behaviour of the modes is inverted, similar to the Schwarzschild-AdS black hole and RNdS. Interestingly enough, for small values of the angular number the dominant near extremal mode becomes purely imaginary (just damping), while that for higher-values of the angular number the dominant modes present a frequency of oscillation.

\acknowledgments
  Y.V. acknowledge support by the Direcci\'on de Investigaci\'on y Desarrollo de la Universidad de La Serena, Grant No. PR18142. \\

\bibliography{references}




\bibliographystyle{ieeetr}


\end{document}